\documentclass[aps,prd,amsfonts,amsmath,nofootinbib,tightenlines,preprint,showpacs,longbibliography,floatfix]{revtex4-1}
\pdfoutput=1

\usepackage{graphicx}
    \graphicspath{ {./figures/} }
\usepackage{amsmath}
\usepackage{array}
\usepackage[utf8]{inputenc}

\def\bA{\mathbf{A}}
\def\bB{\mathbf{B}}
\def\bX{\mathbf{X}}

\def\tobs{\theta_O}
\def\pobs{\phi_O}
\def\Gammac{\Gamma_{\text{cusp}}}
\DeclareMathOperator{\sign}{sign}

\begin{document}

\title{Gravitational backreaction simulations of simple cosmic string loops}
\author{Jose J. Blanco-Pillado}
\email{josejuan.blanco@ehu.es}
\affiliation{IKERBASQUE, Basque Foundation for Science, 48011, Bilbao, Spain}
\affiliation{Department of Theoretical Physics, UPV/EHU,\\48080, Bilbao, Spain}
\author{Ken D. Olum}
\email{kdo@cosmos.phy.tufts.edu}
\affiliation{Institute of Cosmology, Department of Physics and Astronomy,\\Tufts University, Medford, MA 02155, USA}
\author{Jeremy M. Wachter}
\email{jeremy.wachter@ehu.es}
\affiliation{Department of Theoretical Physics, UPV/EHU,\\48080, Bilbao, Spain}

\begin{abstract}

We present the results of computational gravitational backreaction on simple models of cosmic string loops. These results give us insight into the general behavior of cusps and kinks on loops, in addition to other features of evolution. Kinks are rounded off via an asymmetric and divergent correction to the string direction. The result is that cusps emerge in the place of kinks but the resulting smooth string section has a small amount of energy. Existing cusps persist, but quickly lose strength as backreaction removes energy from the string surrounding the cusp. Both kinks and cusps have their location in space shifted slightly with each oscillation.

\end{abstract}

\maketitle

\section{Introduction}\label{sec:intro}

Cosmic strings may arise in our universe as topological defects in
spontaneous symmetry breaking with a non-simply connected vacuum
manifold \cite{Kibble:1976sj,Vilenkin:2000jqa}.  Although they have
not been detected by any experiment, cosmic strings are a generic
feature of many particle physics models, typically forming after the
inflationary epoch in supersymmetric grand unified field theory
models~\cite{Jeannerot:2003qv}.  They also may result from several
string theory
scenarios~\cite{Sarangi:2002yt,Dvali:2003zj,Copeland:2003bj}.  We
expect cosmic strings to be found either as infinite strings or as
closed, oscillating loops. It is the character of these loops which
will be most important to us in the following work.

Because cosmic strings are massive objects which typically move with
relativistic velocities, they will produce gravitational
waves. Particularly because we are now in the era of
gravitational-wave astronomy, it is very viable to detect (or at least
further constrain) cosmic strings by observations (or
non-observations) of these waves. We might be able to observe a
stochastic gravitational wave background due to the oscillations of
loops, or individual bursts from points on the loops known as
\emph{cusps}, which momentarily develop extremely large Lorentz
factors and thus emit a strong, narrow beam of gravitational
radiation.

The stochastic background and cusp signals are both discussed in the literature~\cite{Vilenkin:1981bx,Hogan:1984is,Vachaspati:1984gt,Accetta:1988bg,Bennett:1990ry,Caldwell:1991jj,Damour:2000wa,Damour:2001bk,Damour:2004kw,Siemens:2006yp,DePies:2007bm,Olmez:2010bi,Sanidas:2012ee,Sanidas:2012tf,Binetruy:2012ze,Kuroyanagi:2012wm,Blanco-Pillado:2013qja,Sousa:2016ggw,Blanco-Pillado:2017oxo,Blanco-Pillado:2017rnf,Cui:2017ufi,Chernoff:2017fll,Ringeval:2017eww,Guedes:2018afo} and sought after in detectors~\cite{others:2016ifn,Herner:2017nle,Abbott:2017mem}. However, the picture is not yet complete. As cosmic strings are massive, extended objects, they are expected to interact with themselves via their own gravitational field --- gravitational backreaction --- which may serve to change the shapes of loops and thus some character of their gravitational spectrum. Investigations into how loops change under backreaction up to this point have been limited by the computational power available at the time~\cite{Quashnock:1990wv} or have used approximations to the effects of backreaction in place of exact calculations on each loop~\cite{Blanco-Pillado:2017oxo}.

In this paper, we present the results of exact calculations of
backreaction on four simple models of loops. In Sec.~\ref{sec:model},
we review cosmic strings, explain our formalism, and demonstrate
how our approach recovers the correct results for some cases in which
the analytic answers are known. In Sec.~\ref{sec:results}, we show the
effects of backreaction for specific loops from each of our models,
and compare with theoretical
predictions~\cite{Blanco-Pillado:2018ael,Chernoff:2018evo}.  In
Sec.~\ref{sec:behavior}, we use our results to make general
predictions and observations about how loop features change in the
presence of backreaction, and predict how these results might apply to
realistic loops and thus the gravitational wave signals we might
observe.  We conclude in Sec.~\ref{sec:conc}.

We work in linearized gravity, which is valid because the
string's coupling to gravity is small. We set $c=1$.

\section{A model of gravitational backreaction on loops}\label{sec:model}

Because the ratio of length to thickness of a cosmic string is
typically of order $10^{40}$ or more, it is a good approximation to
treat it as a as one-dimensional object. Thus, a string sweeps out a
worldsheet in spacetime, and its motion can be described by a timelike
$\tau$ and a spacelike $\sigma$ parameter. As usual, we will choose
these parameters so that the metric on the worldsheet is conformally
flat, $\gamma_{\tau\sigma} = 0, \gamma_{\tau\tau} =
-\gamma_{\sigma\sigma}$.  In that case, the general solution to the
Nambu-Goto equations of motion describing the position of the string's
worldsheet in flat space is
\begin{equation}\label{eqn:xab}
    X (\sigma, \tau) = \frac{1}{2} \Big[ A(\tau-\sigma) + B(\tau+\sigma)\Big]
\end{equation}
where $A$ and $B$ are 4-vector functions whose tangent vectors $A'$
and $B'$ are null.  We may also use the null parameterization
$u=\tau+\sigma$ and $v=\tau-\sigma$, which we will choose for the
majority of this work.

In flat space, we can further choose the timelike parameter to be the
coordinate time, $\tau=t$, in which case $\sigma$ parameterizes string
energy (equivalently, the string's \emph{invariant length}), $A'$ and
$B'$ have unit time component, and the corresponding spatial vectors
$\bA'$ and $\bB'$ have unit length.  We may represent $\bA'(v)$ and
$\bB'(u)$ as curves on the unit sphere, which will be useful when we want
to identify cusps and kinks.

A kink is formed whenever there is a discontinuous jump in either $\bA'$ or $\bB'$, which manifests itself as a discontinuous change in direction of the string in space. The discontinuity propagates around the loop at the speed of light.

A cusp is formed by the crossing of the $\bA'$ and $\bB'$ curves;
that is, at a point in spacetime where $\bA'=\bB'$. As a consequence,
at the cusp, $|d\bX/dt| =1$ and $d\bX/d\sigma=0$, and thus the string
doubles back on itself there, and (formally) moves momentarily at the
speed of light.

Now we consider how a string's trajectory is changed by gravitational
backreaction, i.e., the change to the motion of the string due to the
spacetime curvature induced by the stress-energy tensor of the string
itself.  This curvature is always small, being of order $G\mu$, with
$G$ Newton's constant and $\mu$ the string mass per unit length, and
observations limit $G\mu$ to not much more than $10^{-11}$ (e.g., see
\cite{Blanco-Pillado:2017rnf}).  Even though this effect is very
small, it accumulates over many oscillations, and this enables us to
distinguish gauge artifacts, which would oscillate with the changing
metric, from real effects that accumulate over time
\cite{Wachter:2016rwc}.

Thus we consider the string to move in flat space for one
oscillation.  We interpret the changes to the flat-space motion as
an acceleration, given by~\cite{Quashnock:1990wv}
\begin{equation}\label{eqn:acceleration}
    X^\lambda_{,uv} = -\frac 14\Gamma^\lambda_{\alpha\beta}A'^\alpha B'^\beta\,,
\end{equation}
where the $A'$ and $B'$ here are those of the point we are
investigating (the \emph{observation point}), and
$\Gamma^\lambda_{\alpha\beta}$ is the Christoffel symbol there.  In
addition to spatial changes, Eq.~(\ref{eqn:acceleration}) gives a
change to the time components of $A'$ and $B'$.  This disturbs the
choice of $\tau = t$, but we undo this disturbance by
reparameterization, as discussed below in Sec.~\ref{ssec:changes}.

Thus we compute the acceleration from the metric perturbations (and
their derivatives), which we can find by a Green's function integral
over all gravitational sources on the past lightcone of the
observation point. See Refs.~\cite{Wachter:2016rwc,Blanco-Pillado:2018ael} for
details. The corrections to the tangent vectors $A'$ and $B'$ are then
found by integrating the acceleration for one period of oscillation in
the appropriate null direction
\cite{Quashnock:1990wv,Wachter:2016rwc,Blanco-Pillado:2018ael},
\begin{subequations}\label{eqn:delta-a'b'}\begin{align}
    \Delta A'(v) &= 2\int^L_0 X_{,uv}(u,v) du\,,\\
    \Delta B'(u) &= 2\int^L_0 X_{,uv}(u,v) dv\,.
\end{align}\end{subequations}
where $L$ is the invariant length of the string loop.

These corrections to the tangent vectors contain the information about
how $\bA'$ and $\bB'$ move on the unit sphere, as well as how energy
($\sigma$) is lost from each part of the worldsheet. From this
information, we may construct the worldsheet of the backreacted loop.  

Equation~(\ref{eqn:delta-a'b'}) gives the first-order changes to
$A'(v)$ and $B'(u)$, meaning that we accumulate the effect for the
entire oscillation before applying it \cite{Quashnock:1990wv}.
Moreover, since $G\mu$ is so small, we can allow $\Delta A'(v)$ and
$\Delta B'(u)$ to grow for $N\gg1$ oscillations, as long as we keep
$NG\mu\ll1$.  Thus $NG\mu$ is the fundamental parameter in the
simulation; $N$ and $G\mu$ will appear only in this combination.

\subsection{The discretized worldsheet}\label{ssec:discrete}

We expect realistic cosmic string loops to form initially with many
kinks and no cusps, with smooth curves connecting kinks \cite{Blanco-Pillado:2015ana}. These loops may or may not self-intersect, but those which do will quickly (within a single oscillation) reach this self-intersection point and therefore split into two loops. Again, these two child loops may or may not self-intersect, but within a few oscillation times, we expect loops to reach non-self-intersecting trajectories.

We now wish to represent such a loop numerically, so that we can
compute the effect of backreaction on its evolution.  We choose a
representation where $A(v)$ and $B(u)$ are piecewise linear with many
segments, so $A'$ and $B'$ are piecewise constant. We put $N_a$ such
segments in $A$ and $N_b$ in $B$.

This process creates a loop with two kinds of kinks: the true kinks,
which are the discontinuous changes in the string's tangent vectors
seen in the real loops; and the false kinks, which are the
discontinuous changes introduced as a result of discretizing a smooth
curve. When we discuss kinks, the change to kinks, and the location
(or former location) of kinks, we will always mean true kinks.

Now taking our worldsheet functions and assembling the string loop as
in Eq.~(\ref{eqn:xab}), we see that each period of the string
worldsheet is made of $N_aN_b$ patches, each of which is the surface
created by sweeping one segment of $A$ across one segment of
$B$ (or vice versa). These patches we call
\emph{diamonds}.\footnote{Because the segments may be of different
  lengths, these patches are properly parallelograms. Calling them
  diamonds is equal parts history and artistic
  license.}  Consequentially, the edges of these diamonds represent
lines along which $u$ (for an edge parallel to some segment of
$A$) or $v$ (for $B$) are constant, and so all lines
parallel to a diamond's edge are null.\footnote{This relates to the
  earlier point that all kinks move at the speed of light. At the
  edges of the diamonds, the worldsheet jumps between segments of $A$
  or $B$, and thus there are discontinuities there in $\bA'$ or
  $\bB'$.}  For more details on the representation and evolution
of piecewise-linear strings, see
Ref.~\cite{BlancoPillado:2011dq}.

Consider a point on the discretized worldsheet, i.e., inside some diamond. We call this diamond the \emph{observer diamond}. All diamonds which intersect the backward lightcone of this point will be sources of metric perturbations which can contribute to its acceleration. Such diamonds we call the \emph{source diamonds}. The intersection of the past lightcone with the string worldsheet will be a closed line, which is non-self-intersecting if the worldsheet is also non-self-intersecting. We call this closed line the \emph{intersection line}.

We may place restrictions on the intersection line via causality
arguments. First, some terminology. Each diamond has four tips: one at
the largest time coordinate (\emph{future tip}), one at the smallest
time coordinate (\emph{past tip}), and two which are at intermediate
time coordinates (\emph{side tips}) as determined by the segments of $A$
and $B$ that form that diamond. The two diamond edges which
connect the side tips to the future tip are the future edges, and
those connecting side to past are the past edges.

A diamond is a region of a timelike plane.  Such a plane contains two
null directions, and the edges of the diamond lie in these directions.
The intersection of a plane with a cone is a conic section.  Since the
two null directions on the plane are parallel to two lines on the
lightcone, the conic section is a hyperbola, and the intersection line
is a segment of that hyperbola.  The asymptotes of the hyperbola are
parallel to the edges of the diamond, and since we are considering the
past lightcone, the hyperbola opens into the past.  In the observer
diamond, the intersection line is a degenerate hyperbola whose vertex
is the observation point itself.

The hyperbola segment within each diamond is a spacelike path (with
the limiting degenerate case being a null path). Now consider a future
and past edge of a diamond with a common side tip. These edges are
causally connected, and so the hyperbola cannot connect them. Thus,
the intersection line may only cross a diamond in one of four ways:
connecting the two future edges; connecting the two past edges; or one
of the two ways to connect a future edge to its parallel past edge. As
a consequence of this restriction, the intersection line will always
pass through $N_a+N_b$ diamonds, and so we may create an intersection
line for any observation point by considering the causal relationship
of the tips of the worldsheet diamonds to that observation point.
For more details see Ref.~\cite{Wachter:2016rwc}.

\subsection{Changes to the discretized loop}\label{ssec:changes}

Now that we have a discretized loop, we want to find the effect of
gravitational backreaction on it.  To prevent rapid growth in the
amount of data representing the string, we keep it piecewise linear,
with the same pieces as before.  Thus we will compute one $\Delta A'$
for each segment of $A$, and likewise for $B$.  We will choose this
single $\Delta A'$ to be the one computed at the midpoint of the
segment and treat it as representative.  This is accurate providing
that the number of segments is sufficiently large.  To find the
correction to a particular segment of $A$, we will travel in the $u$
direction through the diamonds formed by combining all segments of $B$
with our particular segment of $A$ (and identically with
$A\leftrightarrow B$, $u\leftrightarrow v$).

Our problem reduces to one of finding the corrections along the null lines which bisect each diamond. So, we allow the observation point to move through the observation diamond, and consider how the intersection line changes in response. Because diamonds are timelike surfaces, if a source diamond's future tip is timelike separated from the observation point, the entire source diamond must lie inside the past lightcone of the observation point. So, it cannot be on the intersection line, and cannot contribute to backreaction\footnote{Conversely, if the past tip is spacelike separated, the entire diamond is again outside the past lightcone, and will not contribute.}.

We therefore say that a diamond is ``on the intersection line'' if its
future tip is spacelike separated from the observer, and its past tip
timelike separated. Each future tip is also a past tip of some other
diamond, and so the number of diamonds on the worldsheet is conserved:
as soon as some diamond drops off (due to its future tip now being
timelike separated), some other diamond is added on (due to its past
tip --- the same point --- now being timelike separated). In this way,
we may easily evolve the intersection line as the observation point
evolves by keeping track of where on the observation point's
trajectory it will be null-separated from the future tip of each of
its source diamonds.

\begin{figure}
    \centering
    \includegraphics[scale=0.5]{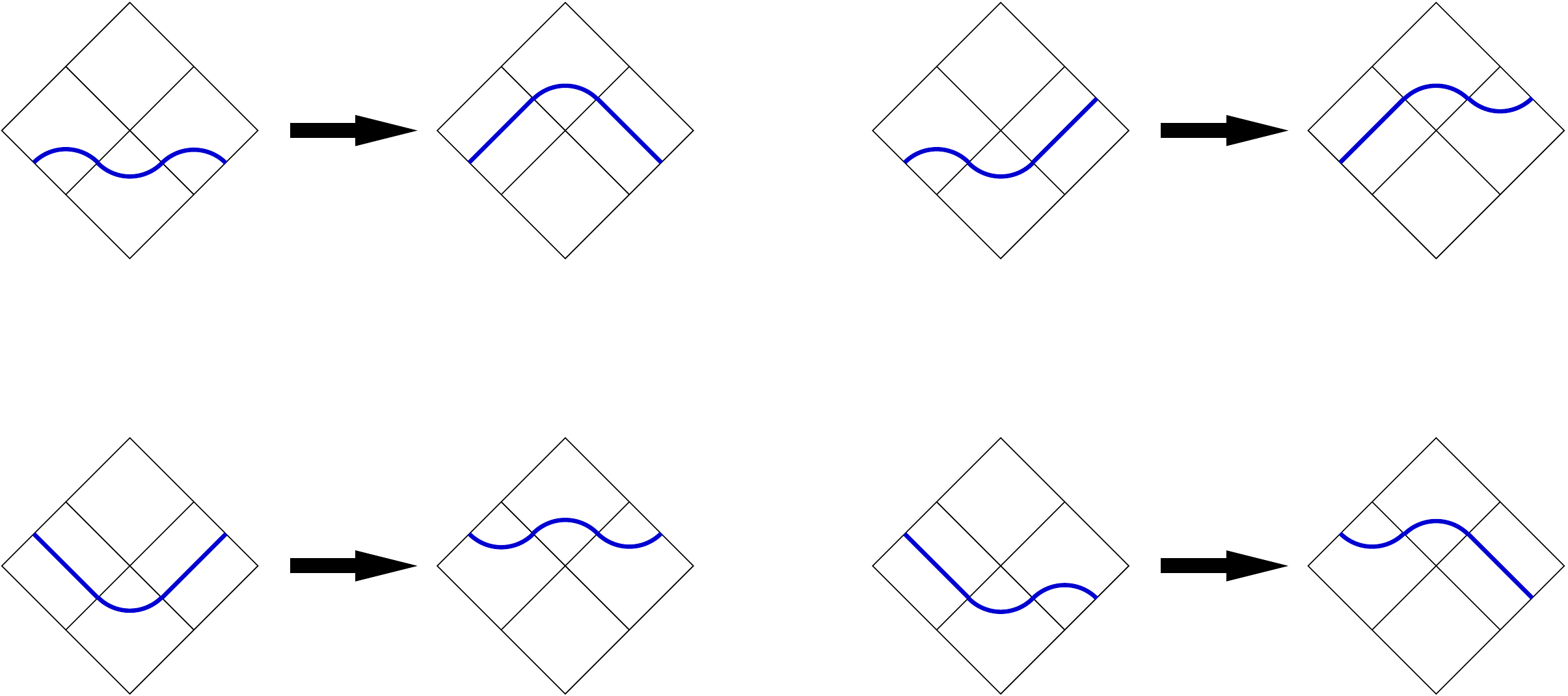}
    \caption{Drawings of how the intersection line (thick blue) must
      change how it crosses the diamonds (bordered in black) as it
      moves forward in time (upward on the page) over some small part
      of the loop worldsheet.}\label{fig:changing-crossings}
\end{figure}

For keeping track of how the crossings of each source diamond change as the observation point moves, we note that the future tip of one diamond is also one of the side points for the two diamonds which border the first diamond on its future edges. So, whenever a source diamond is removed from the intersection line, this is also when the types of crossings for those two neighboring diamonds change. All possible evolutions are shown in Fig.~\ref{fig:changing-crossings}.

We have the general form of the metric perturbation (and its
derivatives) for the four crossing types from
Ref.~\cite{Wachter:2016rwc}, and thus the general form of the
accelerations for each source diamond. These expressions are analytic;
in Appendix~\ref{app:segment-corrections} we integrate them with
respect to $u$ or $v$ to find the contribution of any source diamond
to the tangent vector correction from the parameters of the
source and observer diamonds, plus the range in $u$ or $v$ along the
observer diamond's line of motion for which the source diamond
contributes. The above procedure for evolving the intersection line
gives us all of this information.

Once we have found the tangent vector corrections, we have one more
step before we may construct the new worldsheet. Consider a particular
segment of $B$, with tangent vector
\begin{equation}
    B'^{(0)}=\left(1\,,\,\bB'^{(0)}\right)
\end{equation}
applying over parameter range $\delta\sigma$. The correction, given by
$\Delta B'$, allows us to define a perturbed tangent vector
$B'^{(1)}=B'^{(0)}+\Delta B'$. This vector is still
null to first order, but generally no longer has unit time
component. Consequently, where we previously had $\tau=t$, this is no
longer true (and, similarly, $\sigma$ no longer parameterizes
energy). So, we use the correction to the time component of $B'$ to
reparameterize $\sigma$, defining a new null vector
\begin{equation}\label{eqn:bar-b-1}
    \bar B'^{(1)}=\frac{B'^{(1)}}{1+\Delta B'^t}=\left(1\,,\,\frac{\bB'^{(1)}}{1+\Delta B'^t}\right)\,.
\end{equation}
Whereas before we had $B'^{(0)}=dB^{(0)}/d\sigma$, this new null
vector obeys $\bar B'^{(1)}=dB^{(1)}/d\bar\sigma$, where $\bar\sigma$
is a reparameterization of $\sigma$ which depends on the
correction. The old time component of $B^{(0)}$
ranged over an interval of length $\delta\sigma$, and the new time component of
$B^{(1)}$ ranges over $\delta\bar\sigma$, so
\begin{equation}\label{eqn:dbs}
    \delta\bar\sigma=(1+\Delta B'^t)\delta\sigma\,.
\end{equation}
Usually $\Delta B'^t<0$, so $\delta\bar\sigma<\delta\sigma$,
representing a loss of energy in backreaction.

The spatial part of Eq.~(\ref{eqn:bar-b-1}) gives the new point on the
unit sphere for this segment, and Eq.~(\ref{eqn:dbs}) gives its
new length.  We remove the overbars, and this
reparameterized $\sigma$ once again represents the energy of the loop.
By fixing $t$, we may create snapshots of the loop at a particular
time \cite{Quashnock:1990wv}.

The procedure for any segment of $A$ is analogous, with the note that $A'=-dA/d\sigma$.

\subsection{Testing our approach}\label{ssec:tests}

Before proceeding to our results, let us apply our approach to a
well-studied case and verify that we recover the expected (analytical)
result. To do this, we will consider the Garfinkle-Vachaspati class of
degenerate loops~\cite{Garfinkle:1987yw}, whose worldsheet functions
are lines which go straight out and back in space and have some angle
$\theta$ between them. The resulting loop is planar (and thus
pathological), and the loop frozen at any point in its oscillation is
a rectangle (including the degenerate double line cases).

\begin{figure}
    \centering
    \includegraphics[scale=0.5]{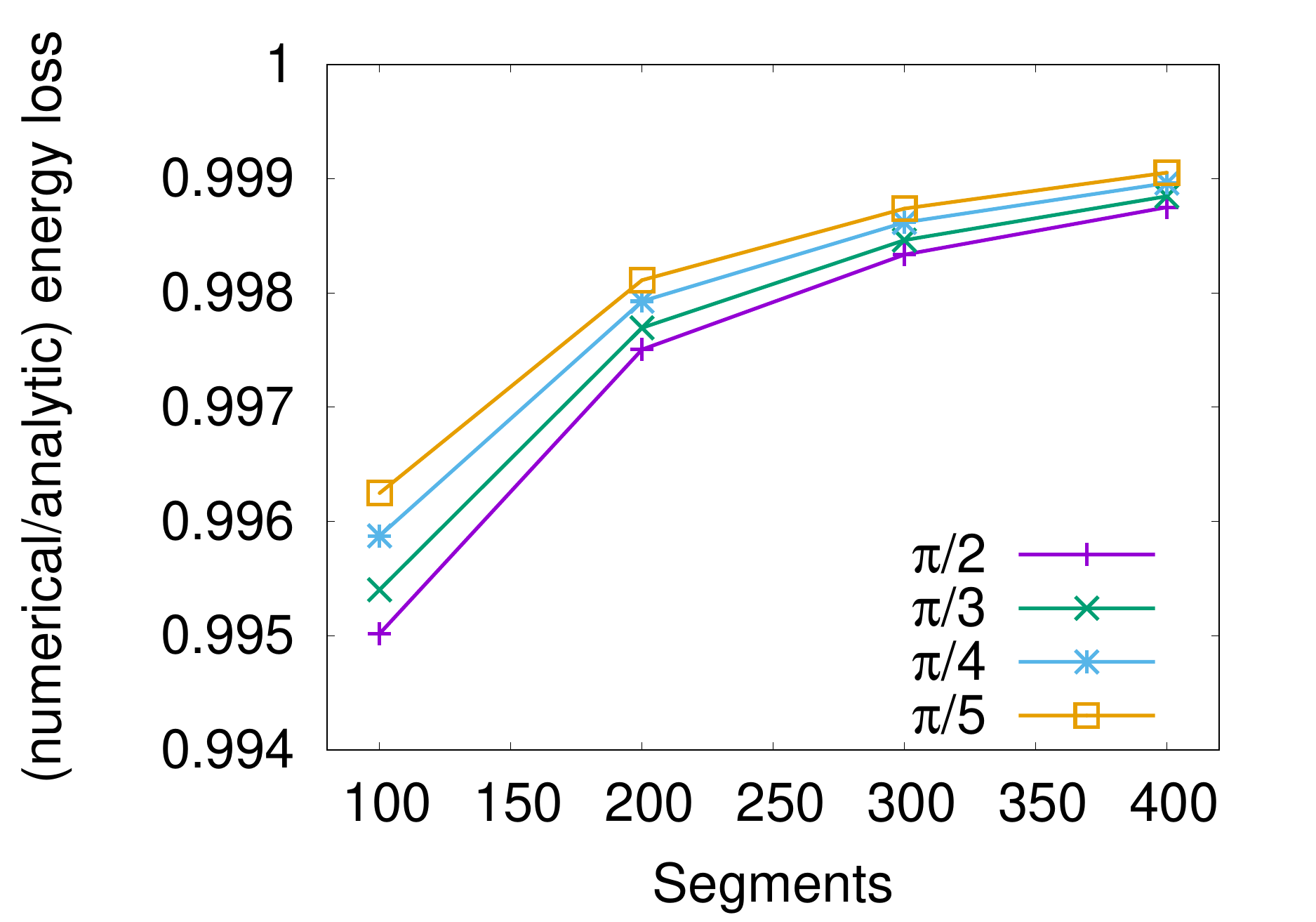}
    \caption{The numerical energy loss to gravitational radiation for
      a single oscillation of the Garfinkle-Vachaspati loop depends on
      the number of segments in the worldsheet functions $A$ and
      $B$. The value reported is the ratio of the numerical result to
      its analytically predicted result. As the number of segments
      increases, the ratio asymptotes towards unity, with the
      numerical result being within $\sim0.1\%$ of the analytical
      result when there are $400$ segments in each of $A$ and $B$. We
      consider four different values for the angle between $A$ and
      $B$, but find that this does not greatly impact the accuracy of
      the result.}\label{fig:test-gamma}
\end{figure}

The energy radiated in one oscillation for these loops can be found by
taking the average gravitational radiation power from
Ref.~\cite{Garfinkle:1987yw} and multiplying by the oscillation
period, $L/2$.  Dividing by the initial loop energy, $\mu L$ gives the
fractional loss of length in one oscillation,
\begin{equation}\label{eqn:gv-length-loss}
    \frac{\Delta L}{L} = \frac{16G\mu}{\sin^2\theta}\left[(1+\cos\theta)\ln\left(\frac{2}{1+\cos\theta}\right)+(1-\cos\theta)\ln\left(\frac{2}{1-\cos\theta}\right)\right]\,.
\end{equation}
For our test, we will perform gravitational backreaction on the Garfinkle-Vachaspati loops with $\theta=\{\pi/2,\pi/3,\pi/4,\pi/5\}$, and at each $\theta$ for $\{100,200,300,400\}$ segments in each of $A$ and $B$, for a single oscillation. Then we will determine the fraction of length lost by these loops using the numerical code that computes the backreaction and divide them by the expected analytic value. The plots of these results are in Fig.~\ref{fig:test-gamma}. Because our code accumulates the effect of backreaction over one oscillation before changing the loop, the loss of length it predicts for a pristine Garfinkle-Vachaspati loop should be exactly that of Eq.~(\ref{eqn:gv-length-loss}). We can therefore compare the numerical result, from the first oscillation only, to the analytic prediction above.

In all cases, we see that as the number of segments increases, the accuracy of our calculation improves. By the time we are at $400$ segments in each of $A$ and $B$, all results are within about $0.1\%$ of the analytic value.

\section{Results for simple models}\label{sec:results}

We now present our results for the simple models which we studied. We
defer detailed interpretation and discussion of these results to
Sec.~\ref{sec:behavior}. 

\subsection{Simulation parameters}

For all the loops studied in this section, we kept certain simulation
parameters constant.  First, we discretized all loops with $400$
segments in each of $A$ and $B$, yielding $160,000$ diamonds for the
section of the worldsheet that includes a complete oscillation of a
loop. We gave all segments the same initial length, because
the $A''$ and $B''$ of the loops studied have almost (for $B$, exactly)
the same magnitude everywhere. In general, when discretizing $A$ and $B$,
one should put in more segments where the rates of change of the
tangent vectors are higher, but this is not a concern until they are
much higher at one place than at another.

Second, we evolved all loops for $200$ iterations with a step of
$NG\mu=10^{-4}$ per iteration. These values were chosen so that by the
end of the evolution, the loops would be roughly halfway dissipated,
per the following. The energy lost per oscillation of the loop to
gravitational radiation is \cite{Vilenkin:2000jqa}
\begin{equation}
    \Delta E = \Gamma G\mu^2 \left(\frac{L}{2}\right) \,,
\end{equation}
where $\Gamma$ is a dimensionless constant that depends only on the
loop's geometry.  The fraction of the loop which has dissipated after some
number of iterations $n$ is then roughly
\begin{equation}
    f_{\text{diss}} \sim \frac{n \Gamma NG\mu}{2}\,.
\end{equation}
We pick $\Gamma=50$ because the distribution of $\Gamma$ for smoothed
loops from simulations has a strong peak at this
value~\cite{Blanco-Pillado:2015ana}. With the choices above, we then obtain
$f_{\text{diss}}\sim 1/2$, which means that in our simulations we will
follow the loops for roughly half of their lifetimes.  This is only a
rough approximation because the actual $\Gamma$ may not be $50$, we 
have neglected the fact that the loop oscillates more rapidly as it
loses energy, and $\Gamma$ itself will change due to backreaction. This
final effect is discussed in more detail in Sec.~\ref{ssec:changes-gamma}.

Finally, we give all loops an initial length of $2\pi$ in arbitrary length units.

\subsection{Model loops}\label{ssec:simple-models}

Our goal in this paper is to simulate loops with cusps and loops with
kinks (and some with both) to see how these features evolve.  We need
to start with loops that have no self intersections or other
pathological features.

We start with a loop with cusps.  The simplest such loop would be
the 1,1 Burden \cite{Burden:1985md} loop, whose $A$ and $B$ are just
circles.  However, this loop collapses into a double line.  To prevent
that, we perturb $A$ with a third-harmonic term, giving
the Kibble-Turok loop~\cite{Kibble:1982cb}
\begin{subequations}\begin{align}
    \bA'(v) &= \left[(1-\alpha)\cos(v)+\alpha\cos(3v)\right]\,\hat x + \left[(1-\alpha)\sin(v)+\alpha\sin(3v)\right]\,\hat y\nonumber\\
    &\qquad+2\sqrt{\alpha(1-\alpha)}\sin(v)\,\hat z\,,\\
    \bB'(u) &= \cos(u)\,\hat x + \sin(u)\left(\cos\phi\,\hat y + \sin\phi\,\hat z\right)\,,
\end{align}\end{subequations}
where $\alpha\in (0,1)$ gives the magnitude of the perturbation and
$\phi$ sets the angle between the planes of the tangent vectors before
perturbation.\footnote{Another possibility would be the 1,2 Burden
  loop, where $B$ goes around its circle twice.  But this is very
  unlike loops that one would expect to form naturally.}
  This loop has no self intersections, and for
$\alpha<\sin^2(\phi/2)$, it has two cusps.  We will choose the
parameters $\phi=\pi/2$ and $\alpha = 0.1$.  Our results are
qualitatively unchanged by some variation in these values, but for
$\alpha$ much smaller the loop is nearly self intersecting and for
$\alpha$ much larger it has a very different character from the
original Burden loop.

As this is the loop that we will then modify to produce the other
loops, we refer to it as the \emph{canonical Kibble-Turok loop}. Our
motivation for studying this loop is to examine how cusps change as a
result of backreaction.

Any loop with cusps may be converted into one with kinks by
identifying where on the unit sphere the tangent vectors $\bA'$ and
$\bB'$ overlap, then removing some of $\bA'$ and/or $\bB'$ around the
cusp location (and reparameterizing the worldsheet functions so as to
not change the overall length). For each such surgery performed, we
introduce a kink.

We will construct our second loop by removing angle $\pi/2$ in two
places from the path of $\bB'$ in the canonical Kibble-Turok loop, so
that $\bB'$ skips over $\bA'$ instead of intersecting.  Thus we
replace two cusps with two kinks.  (We could vary the amount of
angle removed, but the results are qualitatively similar.)  The
expression for $\bA'$ is unchanged, but now we replace $u$ by $\tilde
u$ in our expression for $\bB'$, where
\begin{equation}
    \tilde u = \left\{\begin{array}{ll}(2u/L)(\pi-\psi)+\psi/2\,, & 0<u\leq L/2\\ (2u/L)(\pi-\psi)+3\psi/2\,, & L/2<u\leq L\end{array}\right.\,.
\end{equation}
We will call this the \emph{broken Kibble-Turok loop}.  Our motivation
for studying it is to examine the ways in which kink evolution under
backreaction differs from cusp evolution under backreaction.

Our third loop is the \emph{twice-broken Kibble-Turok loop}. Now, we
remove wedges of angle $\pi/2$ from both $\bA'$ and $\bB'$ around each
cusp point. Again the tangent vectors no longer intersect on the unit
sphere, but now we have replaced the two cusps with four kinks. Our
motivation for studying this loop is that the scenario in which both
tangent vectors jump over the same point is one generic to realistic
loops. Such structures form from self-intersections, such as when the
loops are produced from long strings or existing loops.

Our fourth and final loop is the \emph{cuspy broken Kibble-Turok
  loop}. Here, $\bA'$ is untouched and $\bB'$ is broken, but the jump
in the latter does not avoid the crossing with $\bA'$. This loop
therefore has two cusps and two kinks. Our motivation for studying it
is to see if the existence of cusps influences the evolution of kinks,
and vice versa.

For brevity's sake, we will refer to these four loops, in the order
presented above, as the canonical, broken, twice-broken, and cuspy
broken loops.

\subsection{Canonical Kibble-Turok results}\label{ssec:canonical}

\begin{figure}
    \centering
    \includegraphics[scale=0.65]{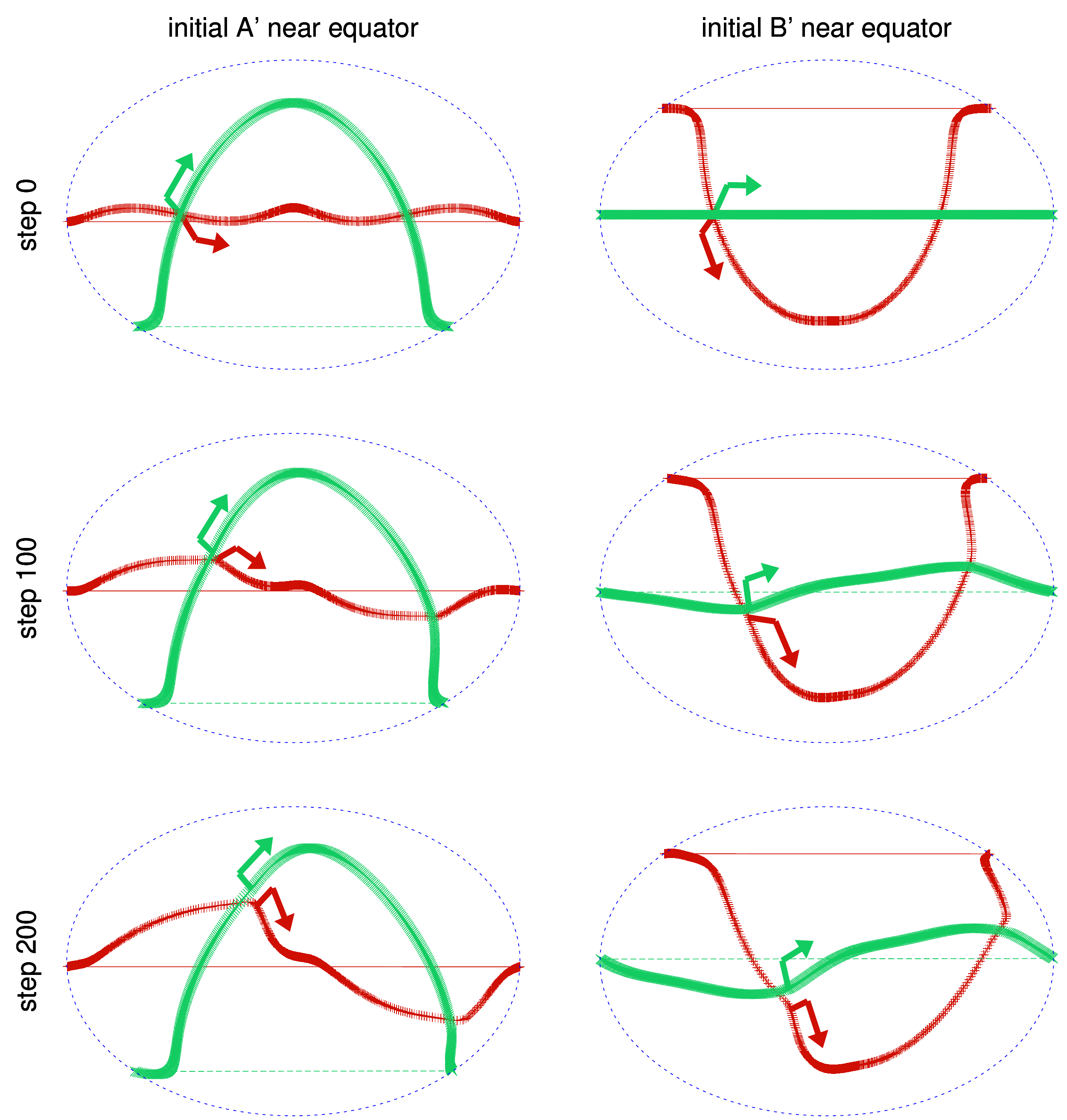}
    \caption{The motion of $\bA'$ (red, $+$) and $\bB'$ (green,
      $\times$) of the canonical loop about the unit sphere as a
      consequence of gravitational backreaction. The rows are,
      top to bottom, $0$, $100$ and $200$ iterations.  In the left
      column we have chosen the projection so that the initial $\bA'$
      lies mostly on the equator, and in the right columns so that the
      initial $\bB'$ lies on the equator.  The three pictures in each
      column use the same projection.  The two pictures in each row
      show the same data with different projections. Arrows show the
      location of segment 0 and the directions in which $u$ and $v$
      increase (i.e., the directions of $\bA''$ and $\bB''$).  Note
      that the cusps are ``dragged'' about the unit
      sphere.}\label{fig:kt-h}
\end{figure}

\begin{figure}
    \centering
    \includegraphics[scale=0.85]{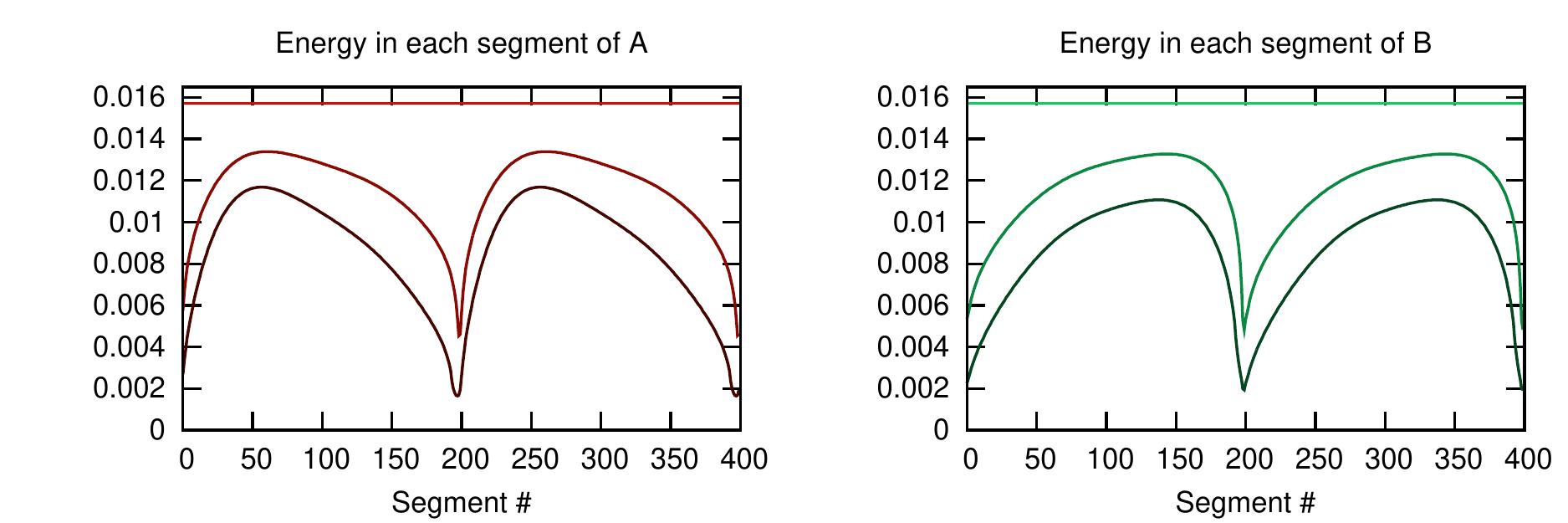}
    \caption{The energy per segment for the canonical loop worldsheet functions changing as a consequence of gravitational backreaction. We have plotted the energies for $0$, $100$ and $200$ iterations, with higher iterations having lower energies/darker colors. The energy loss is preferentially around the cusps and is of the same order on both sides.}\label{fig:kt-s}
\end{figure}

We present the basic results for the canonical loop in Figs.~\ref{fig:kt-h} and \ref{fig:kt-s}. 
Similar plots for the other scenarios will follow in later sections.
To show how the tangent vectors evolve under gravitational
backreaction, we plot $\bA'$ and $\bB'$ on the unit sphere
under the Mollweide projection in Fig.~\ref{fig:kt-h}. The left
panels have $\bA'$ set to lie mostly on the equator, while the
right panels do the same for $\bB'$. This is because the Mollweide
projection is less distorted around the equator, and so using both
projections lets us better understand how each of the tangent vectors
change, and thus how kinks and cusps evolve. We show iterations $0$,
$100$, and $200$, with the last corresponding to a loop of roughly
half its initial length.

The most striking effect we see in Fig.~\ref{fig:kt-h} is that the cusp
locations are dragged about the unit sphere.  The segments of $\bA'$
and $\bB'$ are rotated, primarily in direction of $\bX'' =
(\bA''+\bB'')/2$. This moves the point of the cusp in that direction.
It also moves the individual segments, so that the part of the string
involved in successive cusps changes very little.\footnote{Only the
  small parameter $\alpha = 0.1$ distinguishes the canonical
  Kibble-Turok from the 1,1 Burden~\cite{Burden:1985md} loop.  In that
  loop, the $\bA'$ and $\bB'$ that are equal at the cusp would be
  rotated in exactly the same way, so the part of the string
  contributing to the cusp would be unchanged.}

\begin{figure}
    \centering
    \includegraphics[scale=0.5]{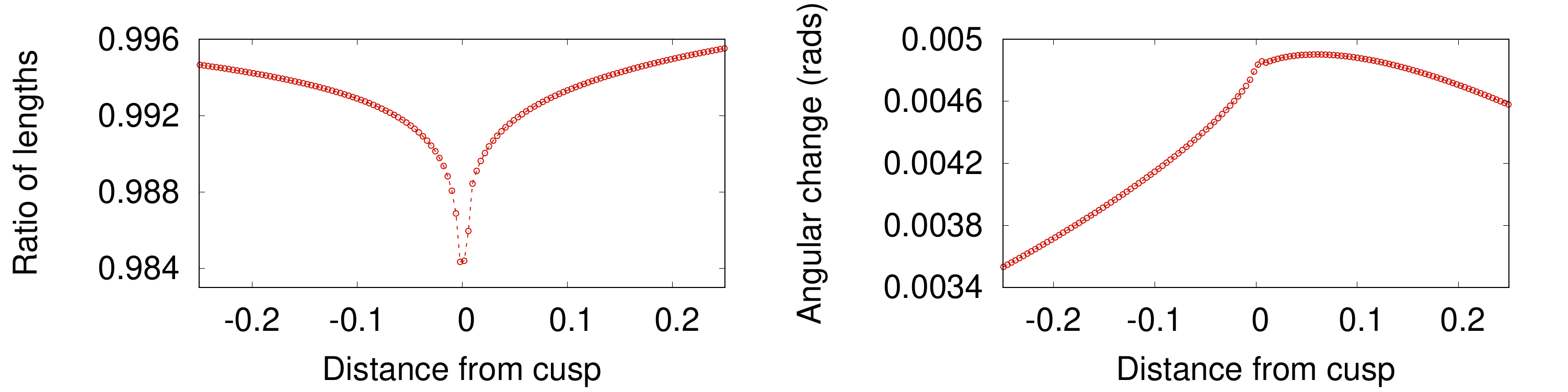}
    \caption{The change of the $2\%$ of overall points in $A$ nearest
      to the cusp due to the first iteration of backreaction. The
      length loss (left) is divergent and mostly symmetric, while the
      angular change (right) is bounded and is greater above the cusp
      (at $v>v_{\text{cusp}}$).}\label{fig:near-cusp-points-a}
\end{figure}

\begin{figure}
    \centering
    \includegraphics[scale=0.67]{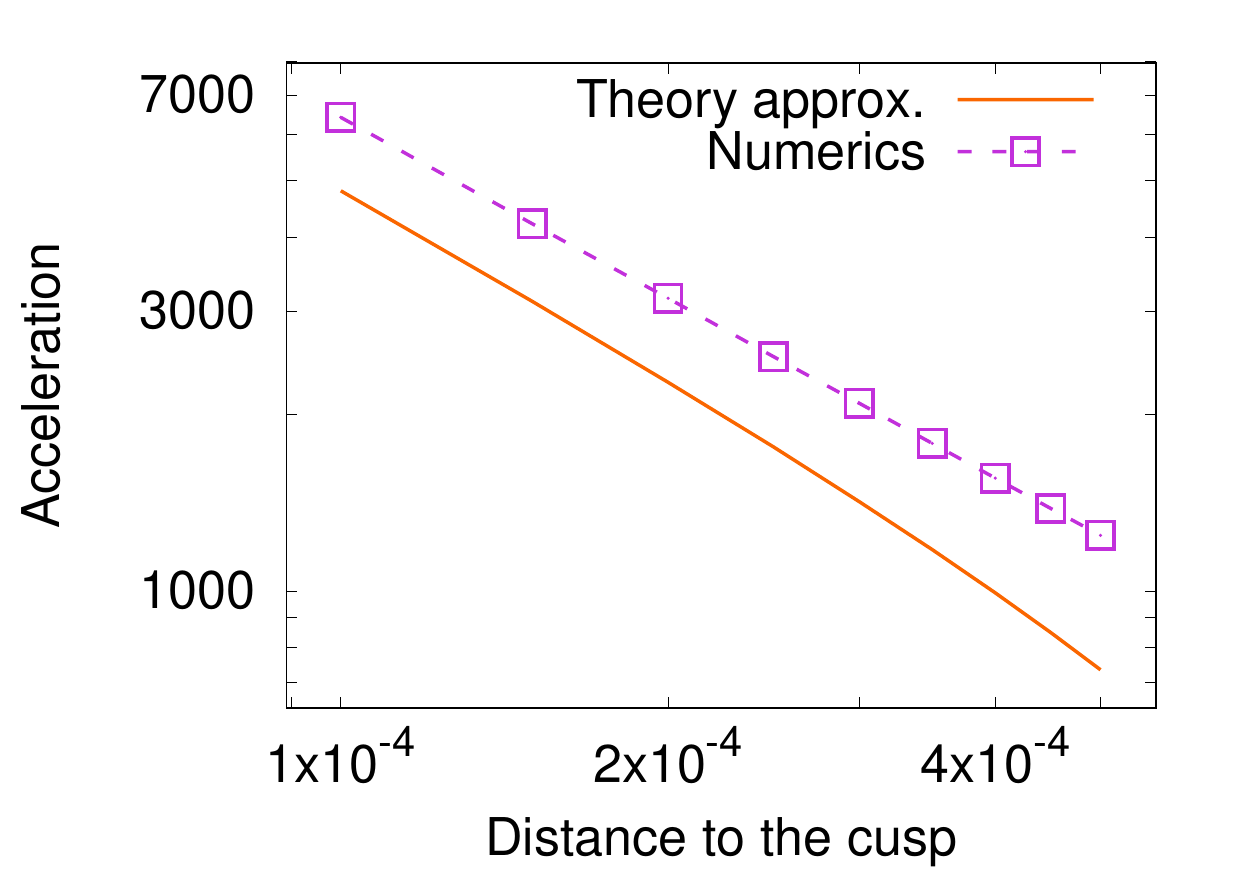}
    \caption{The comparison of numerics to the theoretical
      approximation near the cusp on the canonical loop. The
      theoretical and numerical accelerations both go like the inverse
      of the distance to the cusp. The theoretical approximation gets
      closer and closer to the numerical value as the observation
      point approaches the cusp. For numerical reasons we have not
      been able to get closer than about
      $10^{-4}L$.}\label{fig:cusp-cvm}
\end{figure}

To show how energy is lost due to gravitational backreaction, we plot
the length of all of the segments of both $A$ and $B$ for the same
three iterations as before in Fig.~\ref{fig:kt-s}. This allows us
to see which parts of the string lose more energy during the
backreaction process.  The energy loss is preferentially around the
cusp locations in both $A$ and $B$.  In
Fig.~\ref{fig:near-cusp-points-a},
we look closely at the loss of energy near the cusp in just one
iteration.  It appears to be symmetrical and to diverge as the cusp is
reached.  Refs.~\cite{Blanco-Pillado:2018ael,Chernoff:2018evo}
predicted a logarithmic divergence in the energy emission.  The total
acceleration felt by any worldsheet point which is not at the cusp
goes as the inverse of the distance to the cusp, and integrating along
a line on the worldsheet gives the logarithm.  Following Sec.~V of
Ref.~\cite{Blanco-Pillado:2018ael}, working in the approximation that
we are very near the cusp, we can analytically compute the effect of
each source point, and numerically integrate to obtain the
acceleration on each observation point.  To compare this result to the
acceleration reported by our code, we take a canonical loop and
discretize it to $5\cdot10^5$ diamonds, draw a straight line on the
worldsheet which passes through the cusp, and find the acceleration at
points along this line. This comparison is shown in
Fig.~\ref{fig:cusp-cvm},
where we see that the two approaches are converging up until a
distance from the cusp of $\approx10^{-4}L$.  This is as close as we
can get because of numerical errors, presumably arising from the
high Lorentz factors of the segments near the cusp.

\subsection{Broken Kibble-Turok results}\label{ssec:broken}

\begin{figure}
    \centering
    \includegraphics[scale=0.65]{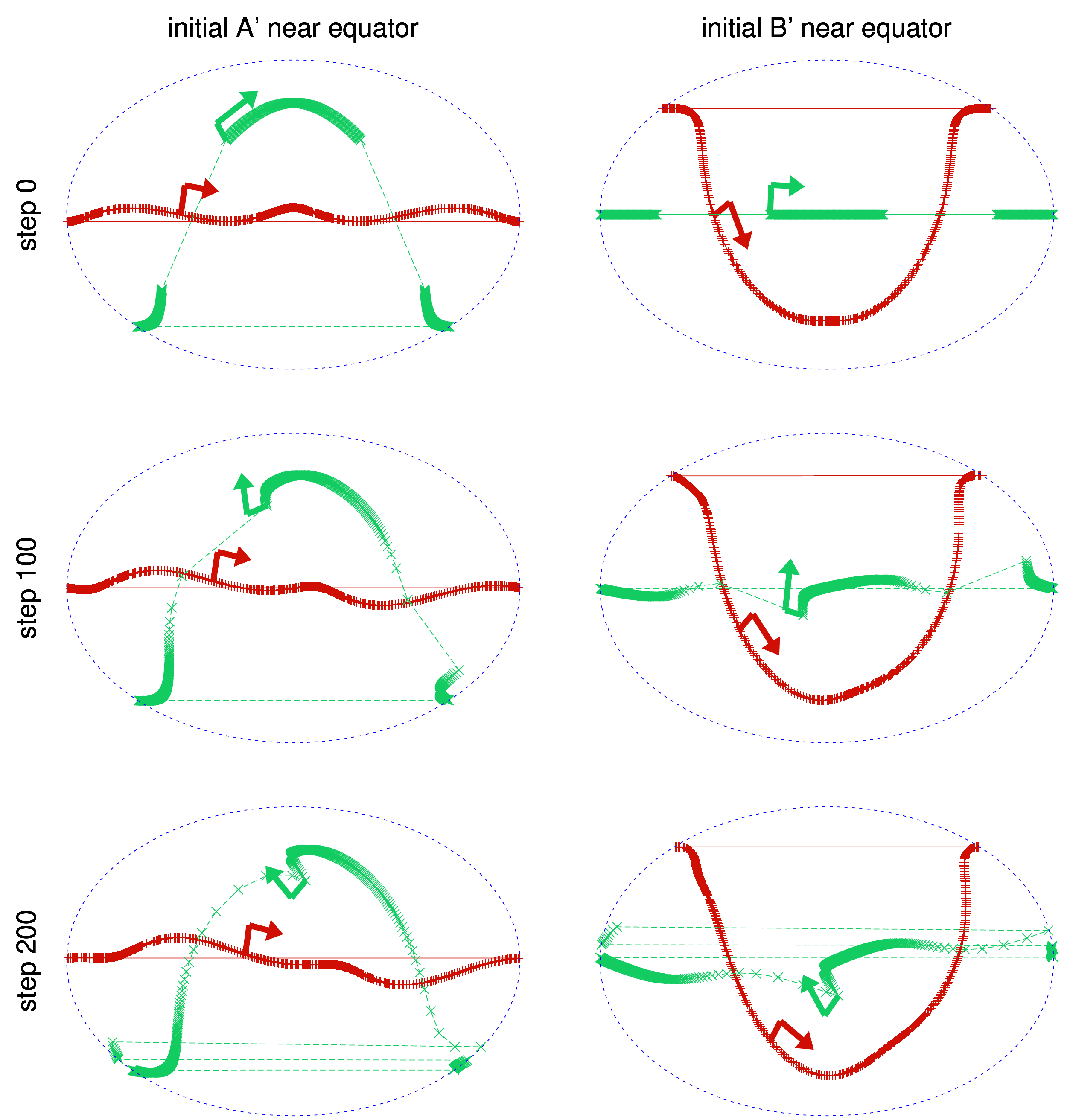}
    \caption{The motion of $\bA'$ (red, $+$) and $\bB'$ (green, $\times$) of the broken loop about the unit sphere.}\label{fig:bkt-h}
\end{figure}

\begin{figure}
    \centering
    \includegraphics[scale=0.85]{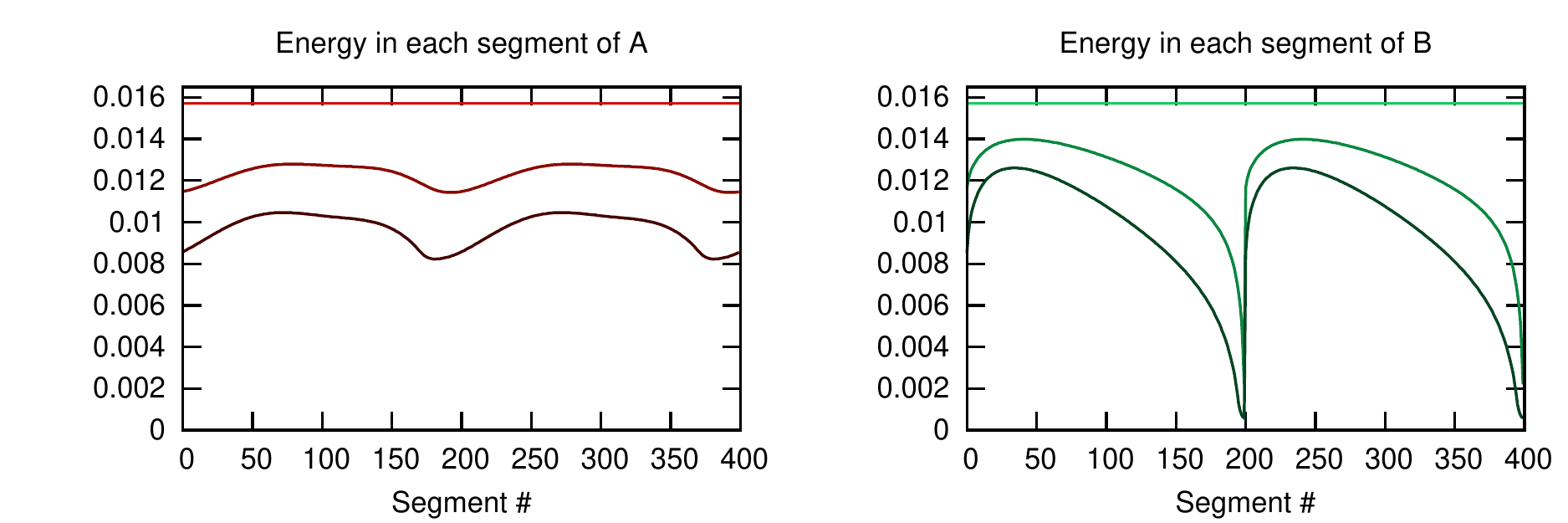}
    \caption{The energy per segment for the broken loop worldsheet functions changing as a consequence of gravitational backreaction. The energy loss is preferentially around the kinks in $B$ and is greater on the side with lower null parameter, while the energy loss in $A$ is less, but happens preferentially around where $B$ jumps over $A$.}\label{fig:bkt-s}
\end{figure}

We present the results for the broken loop in Figs.~\ref{fig:bkt-h}
and \ref{fig:bkt-s}. Now, the cusps have been removed and replaced by
kinks by removing sections of $B$ around the cusps, but keeping the
overall length of $B$ the same.

\begin{figure}
    \centering
    \includegraphics[scale=0.5]{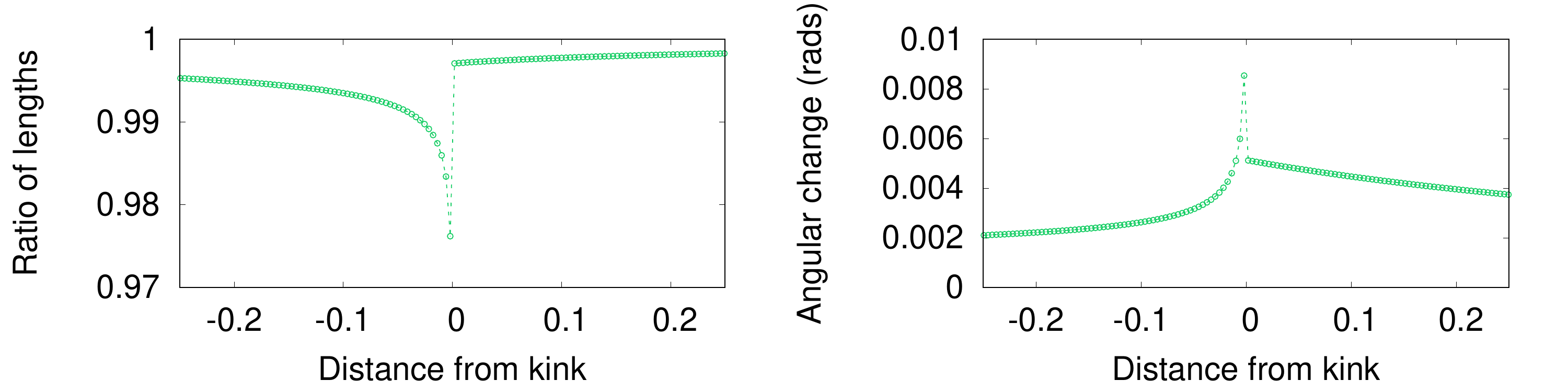}
    \caption{The change of the $2\%$ of overall points in $B$ on both sides of the kink due to one iteration of backreaction. The changes in both length and angle are asymmetric, being divergent below the kink and bounded above.}\label{fig:near-kink-points}
\end{figure}

There is still a preferential loss of energy around the place where
the cusp would be --- the ``jumping-over point'' --- for both $A$ and
$B$. However, it is much more pronounced in $B$. Closer examination
in Fig.~\ref{fig:near-kink-points}
shows that the corrections to the energy and direction of the string
diverge as $u$ approaches the kink position $u_k$ from below, but not
when $u$ approaches $u_k$ from above.  This divergent behavior was
predicted in Refs.~\cite{Blanco-Pillado:2018ael,Chernoff:2018evo}.

\begin{figure}
    \centering
    \includegraphics[scale=0.67]{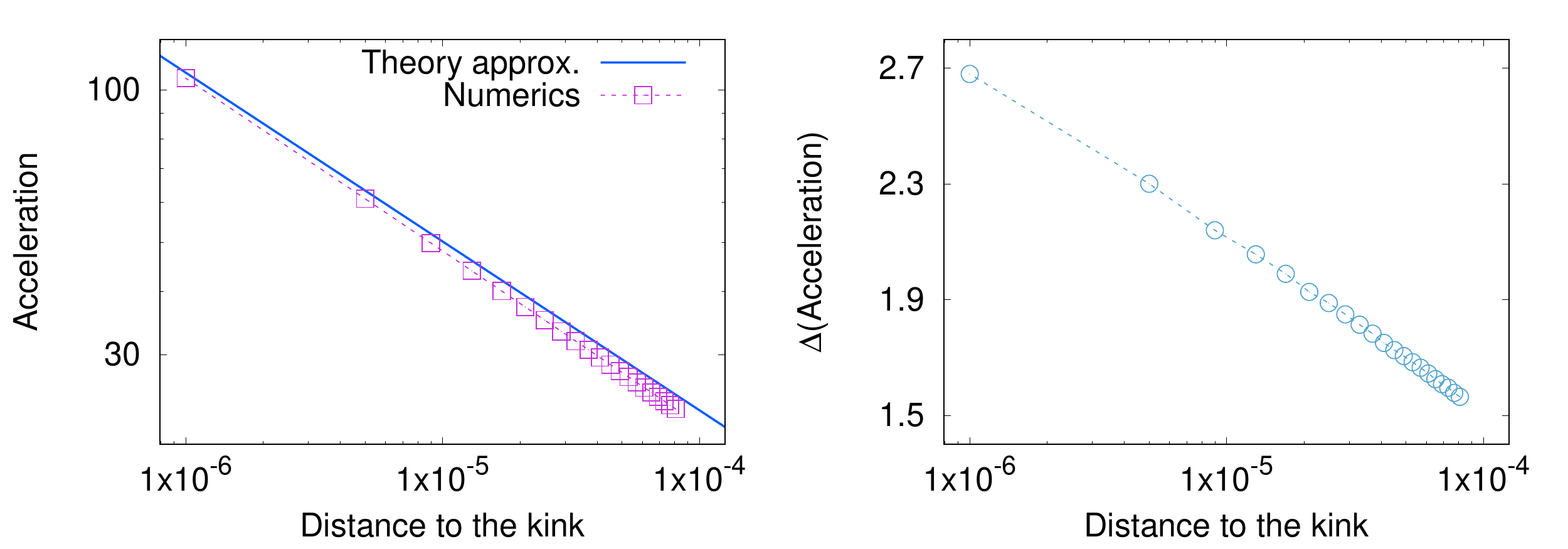}
    \caption{The comparison of numerics to theory for the broken
      loop. Left panel: The theoretical and numerical accelerations
      both go like the inverse cube root of the distance to the
      kink. Right panel: the difference between the theoretical and
      numerical accelerations goes like the logarithm of the distance
      to the kink. This is the lower-order divergent effect predicted
      in Ref.~\cite{Blanco-Pillado:2018ael}.}\label{fig:kink-tvn}
\end{figure}

From Eq.~(71) of Ref.~\cite{Blanco-Pillado:2018ael}, we can calculate
analytically the acceleration felt by a point below a kink.  We work
to leading order as we approach the kink, meaning that we include only
the term that diverges as $(u_k-u)^{-1/3}$ and not a subleading
logarithmic divergence that was predicted but not calculated in
Refs.~\cite{Blanco-Pillado:2018ael,Chernoff:2018evo}.  To compare the
theoretical approximation to the value reported by our code, we take a
broken loop and discretize it to $5\cdot10^5$ diamonds, fix the $v$
index at some arbitrary value, and calculate the transverse
acceleration for varying $u$ index values from a diamond far below the
kink up to the diamond just below the kink. We plot these results in
Fig.~\ref{fig:kink-tvn}.
As the distance to the kink goes to zero, the leading effect in the
numerical acceleration is the predicted $(u_k-u)^{-1/3}$ divergence,
and the coefficient agrees with the theoretical prediction.  The
difference between the numerical and theoretical results shows the
additional effect logarithmic in the distance to the kink.

Is it possible that this divergence could be an artifact, rather than
a real, physical effect?  First of all, it cannot be an artifact of the
choice of spacetime gauge, because the perturbation of the spacetime
metric around Minkowski space is always of order $G\mu$ and does not
grow with time, while the effect of the gravitational backreaction on
the string shape has secular growth.  It also is not an artifact of
the worldsheet gauge, i.e. the choice of the parameters $\tau$ and
$\sigma$.  These can be chosen to satisfy the conformal gauge
conditions, thus defining the functions $A(v)$ and $B(u)$ and their
tangent vectors $A'$ and $B'$, from which we see that $B'$ has rapid
variation over a small range of $u$.  Finally, the rapid change in
$B'$ is not an artifact of which $B'$ at later times we compare with
which $B'$ at earlier times.  The directions in which $B'$ points after
backreaction are novel: no element of $B'$ pointed close to these
directions before, so we can see that these changes are large in a
real sense and do not depend on any gauge or parameter choices.

How does this divergent correction arise? While there is no divergence
in the metric perturbation at a point near the kink, there is a
divergence in the derivatives in the null direction of the kink's
propagation (i.e., in $v$ for a kink in $B$). This divergence is as
the inverse cube root of the null distance from the observation point
to the kink (i.e., as $(u_k-u)^{-1/3}$ for a kink in $B$), which
explains why we see a divergence in $B$ but not $A$ for the broken
case.  The divergence is integrable, so when we integrate the
acceleration with respect to $u$ to find $\Delta A'$, it becomes a
finite correction. But to find $\Delta B'$ we integrate with respect
to $v$. Instead of crossing the kink as we integrate and thus removing
the divergence, we travel around the worldsheet parallel to the kink,
and so the divergence persists. However, as with the cusps, the total
correction to the worldsheet ($\Delta A$ or $\Delta B$) will always be
non-divergent regardless of which worldsheet function contains the
kink, as finding these corrections requires integrating with respect
to both $u$ and $v$.

As discussed in Ref.~\cite{Blanco-Pillado:2018ael}, the correction to
$B$ diverges as one approaches the kink from one side. The kink will
be rounded off by backreaction, and so we expect cusps to form. This
is not the same, however, as the cusps which form due to the toy model
of backreaction of Ref.~\cite{Blanco-Pillado:2015ana}. There, the
authors smoothed the string by convolving the functions $A'$ and $B'$
with a Lorentzian,\footnote{The reason for this choice and the details
  of the implementation are discussed in
  Ref.~\cite{Blanco-Pillado:2015ana}.} which replaces a sharp kink by
a smooth curve. For the rounding off discussed here, as can be seen in
Figs.~\ref{fig:bkt-h} and \ref{fig:bkt-s}, the curvature happens over
a very short amount of length (due to the energy near the kink being
preferentially depleted). The effect seen here leads to a much higher
$A''$ or $B''$ over a much shorter range compared to the convolution
procedure of Ref.~\cite{Blanco-Pillado:2015ana}. The cusps which
actually form due to backreaction will therefore be weaker than
previously predicted.

In the canonical loop, we expect the preferential loss of energy around the cusp to lead to the weakening of cusps. For the broken loop, we anticipate backreaction to lead to the formation of cusps, but because the string has already lost a good amount of energy modifying the kinks, the created cusps will be very weak.

\subsection{Twice-broken Kibble-Turok results}\label{ssec:twice-broken}

\begin{figure}
    \centering
    \includegraphics[scale=0.65]{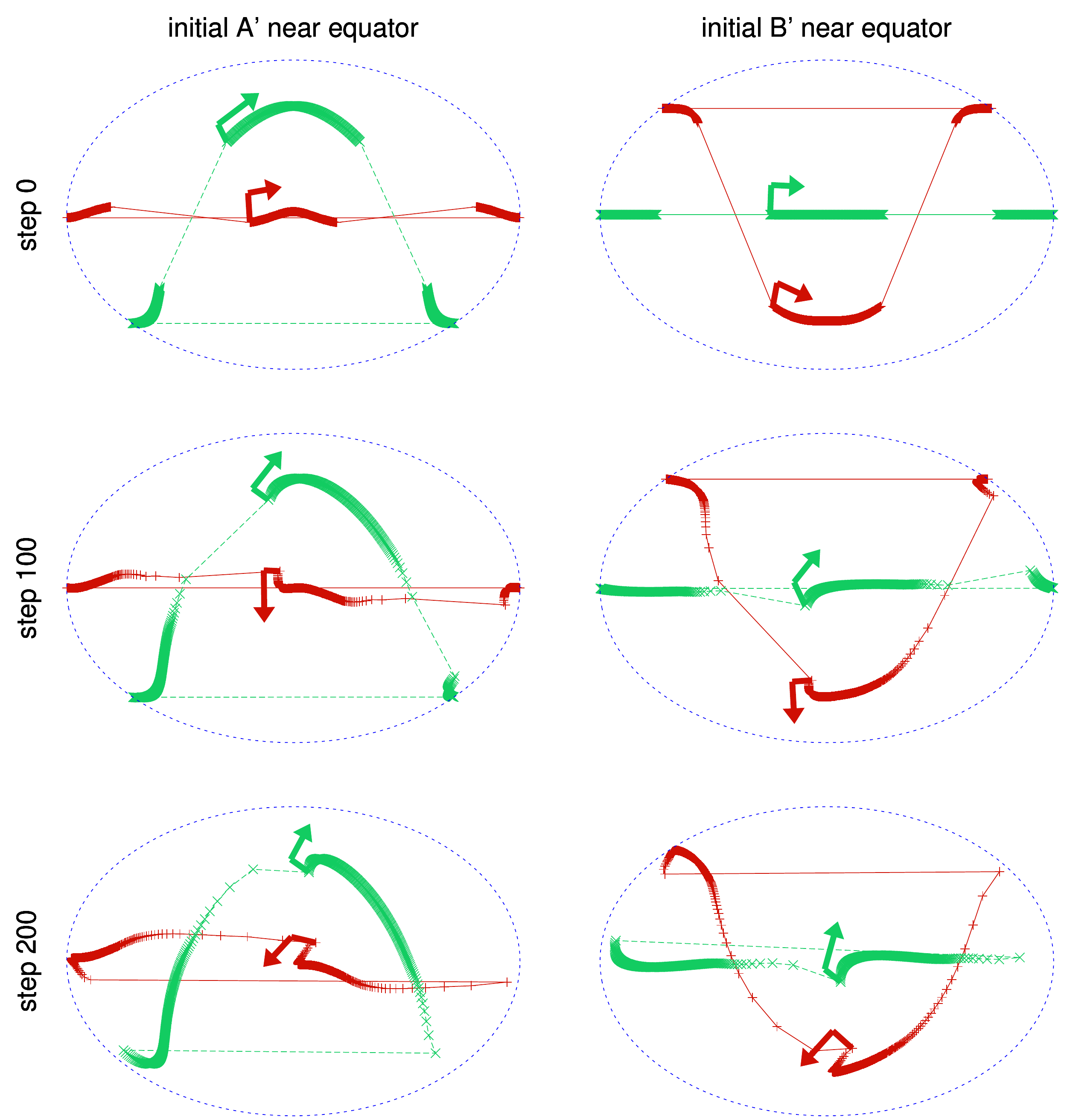}
    \caption{The motion of $\bA'$ (red, $+$) and $\bB'$ (green, $\times$) of the twice-broken loop about the unit sphere.}\label{fig:2bkt-h}
\end{figure}

\begin{figure}
    \centering
    \includegraphics[scale=0.85]{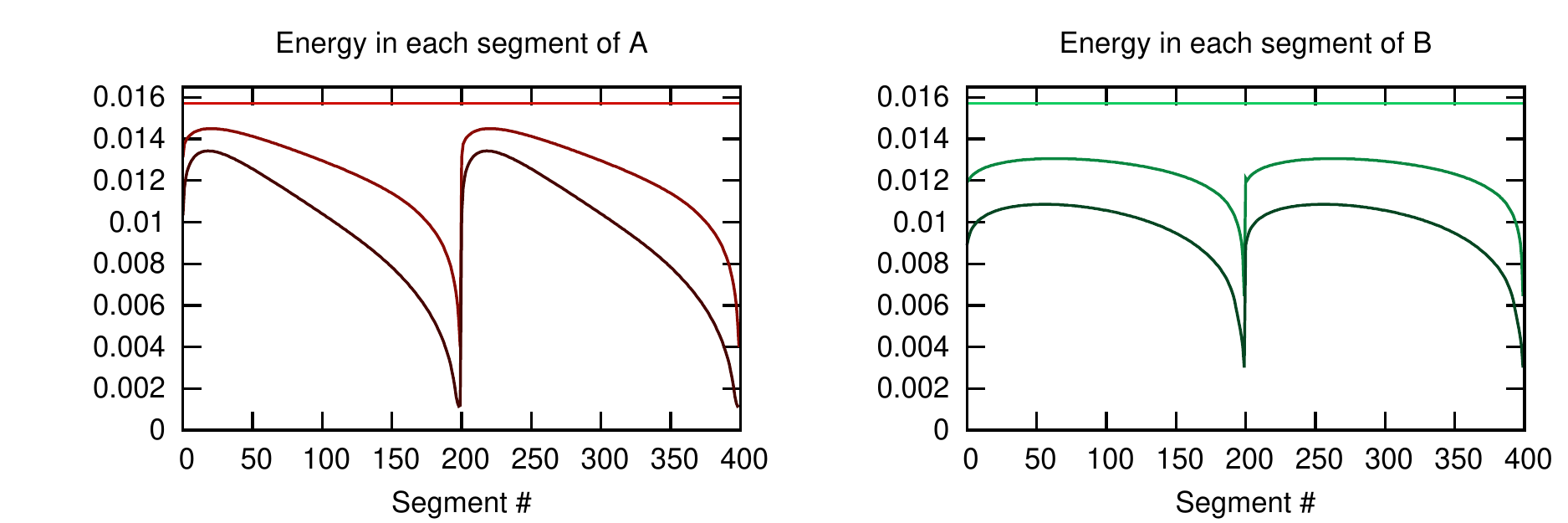}
    \caption{The energy per segment for the twice-broken loop worldsheet functions changing as a consequence of gravitational backreaction. The energy loss is preferentially around the kinks and is greater on the side with lower null parameter.}\label{fig:2bkt-s}
\end{figure}

We present the results for the twice-broken loop in Figs.~\ref{fig:2bkt-h} and \ref{fig:2bkt-s}.
As with the broken loop, we see a preferential loss of energy for the
segments around the kinks, although now it affects both $A$ and $B$
because both contain kinks. We moreover see the same change to the
kinks as observed in the singly-broken case, both in rounding and in
dragging.

\subsection{Cuspy broken Kibble-Turok results}\label{ssec:cuspy broken}

\begin{figure}
    \centering
    \includegraphics[scale=0.65]{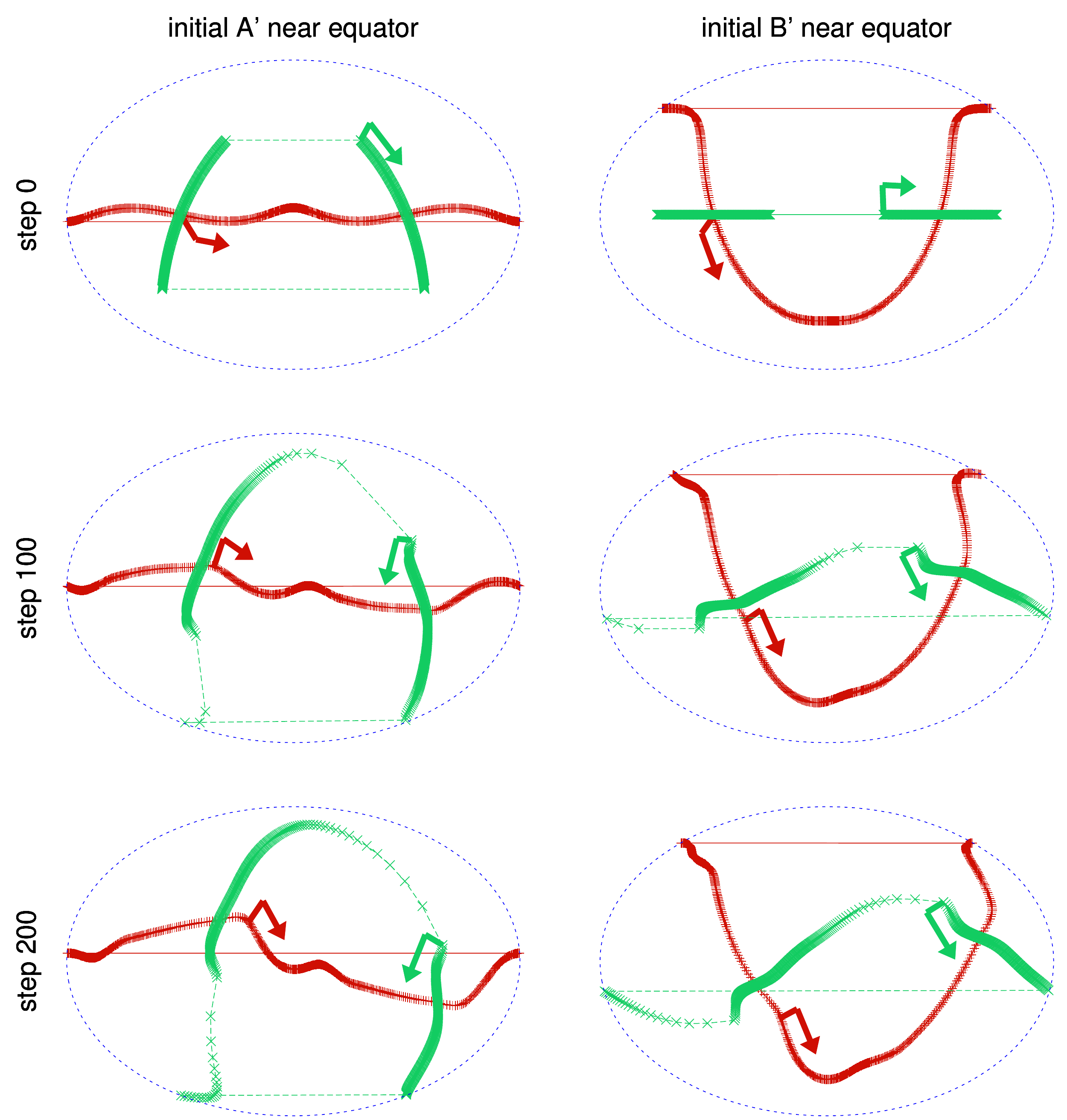}
    \caption{The motion of $\bA'$ (red, $+$) and $\bB'$ (green, $\times$) of the cuspy broken loop about the unit sphere.}\label{fig:mkt-h}
\end{figure}

\begin{figure}
    \centering
    \includegraphics[scale=0.85]{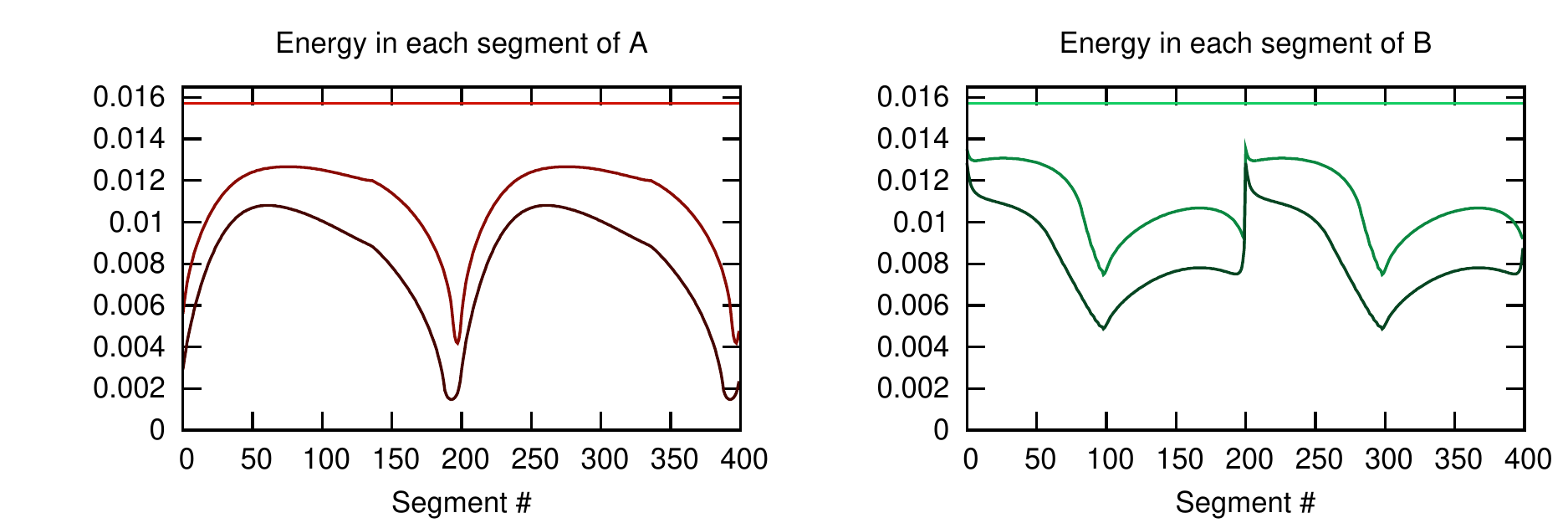}
    \caption{The energy per segment for the cuspy broken loop worldsheet functions changing as a consequence of gravitational backreaction. The energy loss is preferentially around the kinks and cusps; in the former case, the loss is greater on the side which has lower null parameter. The cusps are near segment $200$ and $400$ for $A$, and $100$ and $300$ for $B$. The kinks are shifted by $100$ segments from the cusps for both $A$ and $B$.}\label{fig:mkt-s}
\end{figure}

Lastly, we present the results for the cuspy broken loop in Figs.~\ref{fig:mkt-h} and \ref{fig:mkt-s}. 
Now, we see that the preferential loss of energy at the kinks and cusps happens at slightly different rates, and with quite different behaviors. The depletion of energy and curving of the string near the kink happens preferentially on one side, whereas the depletion and curving near the cusp happen with roughly the same magnitude on both sides (although the process is not symmetric). Both the cusps and the kinks are dragged around the unit sphere, with the kinks being dragged faster.

The changes to the cusps and kinks in the cuspy broken case appear to
be non-interfering, or at most weakly interfering. By this we mean
that the cuspy broken loop behaves more or less like the superposition
of the canonical and broken loops. This suggests that cusps and kinks
only significantly change parts of the string very close to them, and
also that the evolution of the kink does not depend strongly on
whether or not it is avoiding a cusp.

\section{General behavior under backreaction}\label{sec:behavior}

\subsection{Changes to cusps}\label{ssec:change-cusps}

The locations of the cusps on the unit sphere are changed, in a
process we have referred to as \emph{dragging}. So, each time the cusp
reappears, the direction in which it points is slightly
different. This behavior was noted in Ref.~\cite{Quashnock:1990wv},
where the cusps were said to be \emph{delayed}. For example, in
Fig.~\ref{fig:kt-h} we can clearly see that $\bA'$ is dragged in the
direction of $\bB''$, and looking closer we see that $\bB'$ is dragged
in the direction of $\bA''$ also.  When we decide to describe the
string in terms of $A$ and $B$ there is an arbitrary choice of the
direction of $\sigma$, which determines which is $A$ and which is $B$,
and similarly which is $u$ and which is $v$.  It does not, however,
affect the direction of advance of $u$ and $v$, because this is always
toward the future.  Thus the directions of the derivatives of $A$ and
$B$ are not reversed.  Since the dragging effect is symmetrical under
exchange of $A$ and $B$, it does not depend on the choice of direction
of $\sigma$.

\begin{figure}
    \centering
    \includegraphics[scale=0.67]{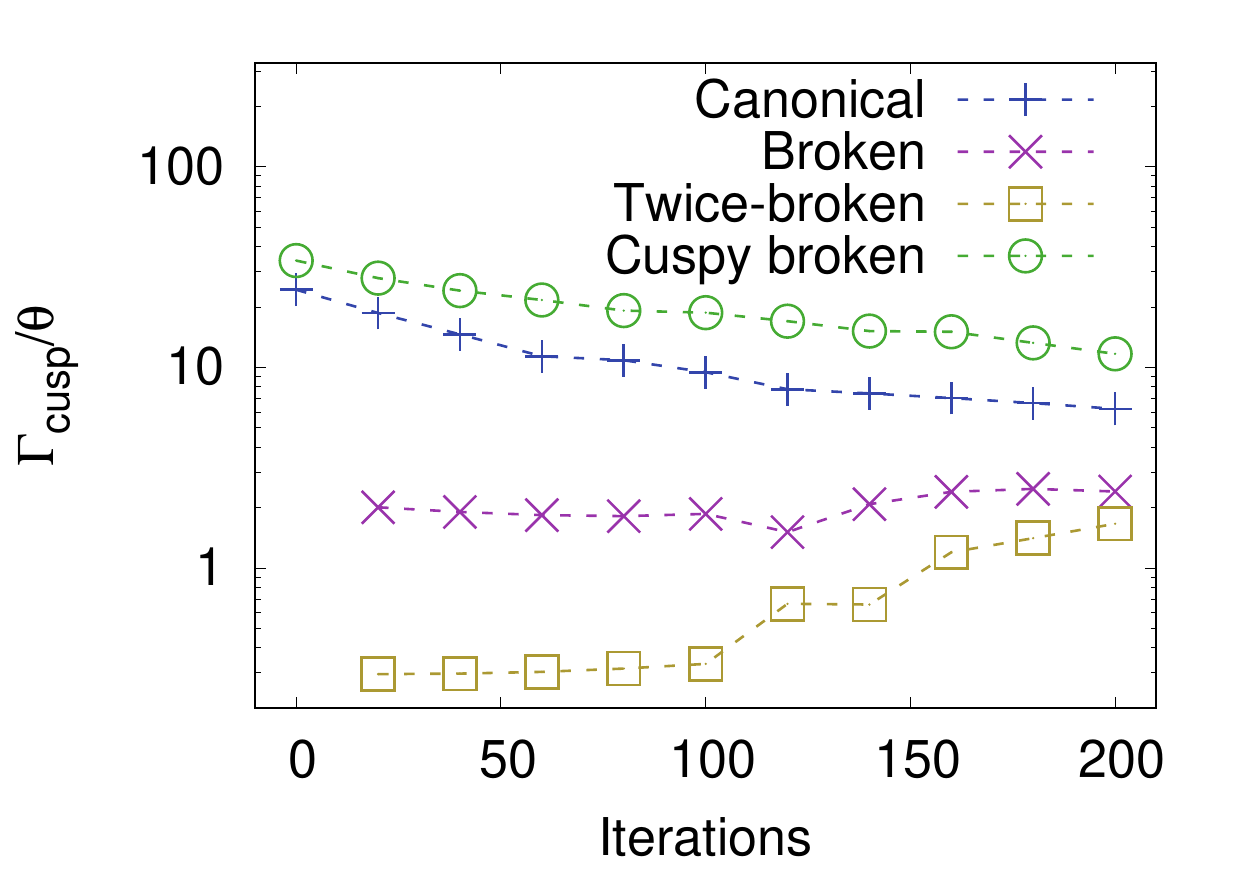}
    \caption{The changes in cusp strength (see text)
     due to gravitational backreaction.}\label{fig:cusp-strength}
\end{figure}

The energy removed from the string by backreaction is preferentially
taken from the string around the cusps, which leads to the cusps
becoming weaker.  The angular power density due to a cusp diverges as
one approaches the cusp direction, but the total power radiated is
finite.  We model the cusp by expanding the string near the cusp in a
Taylor series and following Appendix A of
Ref.~\cite{Blanco-Pillado:2017oxo}.  Let us define $\Gammac(\theta)$
as the contribution to $\Gamma$ coming from radiation into the cone of
directions within angle $\theta\ll 1$ of the cusp direction.  We
compute this quantity in Appendix~\ref{app:gamma-cusp} below.  We find that
$\Gammac(\theta)$ is proportional to $\theta$, so
$\Gammac(\theta)/\theta$ does not depend on $\theta$ and characterizes
the strength of the cusp.  In Fig.~\ref{fig:cusp-strength},
we plot this strength as a function of
the amount of backreaction.

Our measure, $\Gammac(\theta)/\theta$, is based on
average power due to the cusp, not the energy of each burst.  If a
loop were to shrink without changing shape, this quantity would be
constant.  We found this measure useful for understanding changes in
loop shape, but if you are interested in the observability of bursts,
you should multiply by $L/2$ to get the burst energy per unit
$\theta$.  That measure would see an additional drop in energy due to the
shrinkage of the loop.

Cusps which are initially present weaken over time, with the
contribution to $\Gamma$ after iteration 200 being roughly half of
what it was initially.  The cuspy broken loop has a stronger cusp than
the canonical loop, but this is due to how we constructed the
loops. Recall that all loops start with the same length. Thus the
wedge removed from the cuspy broken loop's $B'$ means that the same
amount of energy as in the canonical case is spread across less
angular distance, and thus the $\bB''$ at the cusp is smaller in
the cuspy broken case, so the radiation is stronger.

Cusps that develop on loops which lacked them initially start out weak
and never grow as strong as the cusps that were there from the
beginning.  In the case of the broken loop, backreaction on the kink
produces a somewhat smooth segment of $\bB'$ that crosses the
pre-existing smooth $\bA'$ to form a cusp.  In the case of the
twice-broken loop, both $\bA'$ and $\bB'$ start with kinks.
Thus in this case there is initially much less string involved in the
cusp (i.e., both $\bA''$ and $\bB''$ are much larger), and thus the
cusp radiation is much weaker than in the singly-broken case.

Since the weak cusps are getting stronger and the strong cusps weaker,
there may be a convergence to a single strength of cusps in all cases,
but it is hard to tell.  Even if so, such a thing would happen only
after most of the loop's energy has been lost.

\subsection{Changes to kinks}\label{ssec:change-kinks}

The locations of the kinks on the unit sphere are dragged, again in
the general direction of $\bX''$. As far as we know this behavior has
not been discussed before.  We cannot comment extensively on the
relative rates of cusp and kink dragging, but they appear to differ by
less than an order of magnitude.

The energy removed from the string by backreaction is also
preferentially taken from the string around the kinks --- more
strongly for whichever of $A$ or $B$ contains the discontinuity, but both
are affected. While the cusps lose energy roughly equally on both
sides, the kinks lose energy in a very asymmetric fashion, with the
side above the kink being almost unaffected and the side below being
quickly depleted \cite{Blanco-Pillado:2018ael}.

The kink is rounded off, also in an asymmetrical fashion, as we see,
for example, in Fig.~\ref{fig:bkt-h}.  In the bottom panels, several
of the original segments now fill the gap seen in the top panels.
However, we also see from Fig.~\ref{fig:bkt-s} that little of the
energy associated with each of these segments still remains.  Thus
backreaction replaces the kink by a curved section, but this curvature
is confined to a quite small region of the string. While the kink
has been rounded off, and so is no longer completely preventing cusps,
any cusps which do form will be weak compared to the cusps we studied
which were present at a loop's creation. We can see this behavior in
Fig.~\ref{fig:cusp-strength}.

\subsection{Changes in $\Gamma$}\label{ssec:changes-gamma}

\begin{figure}
    \centering
    \includegraphics[scale=0.67]{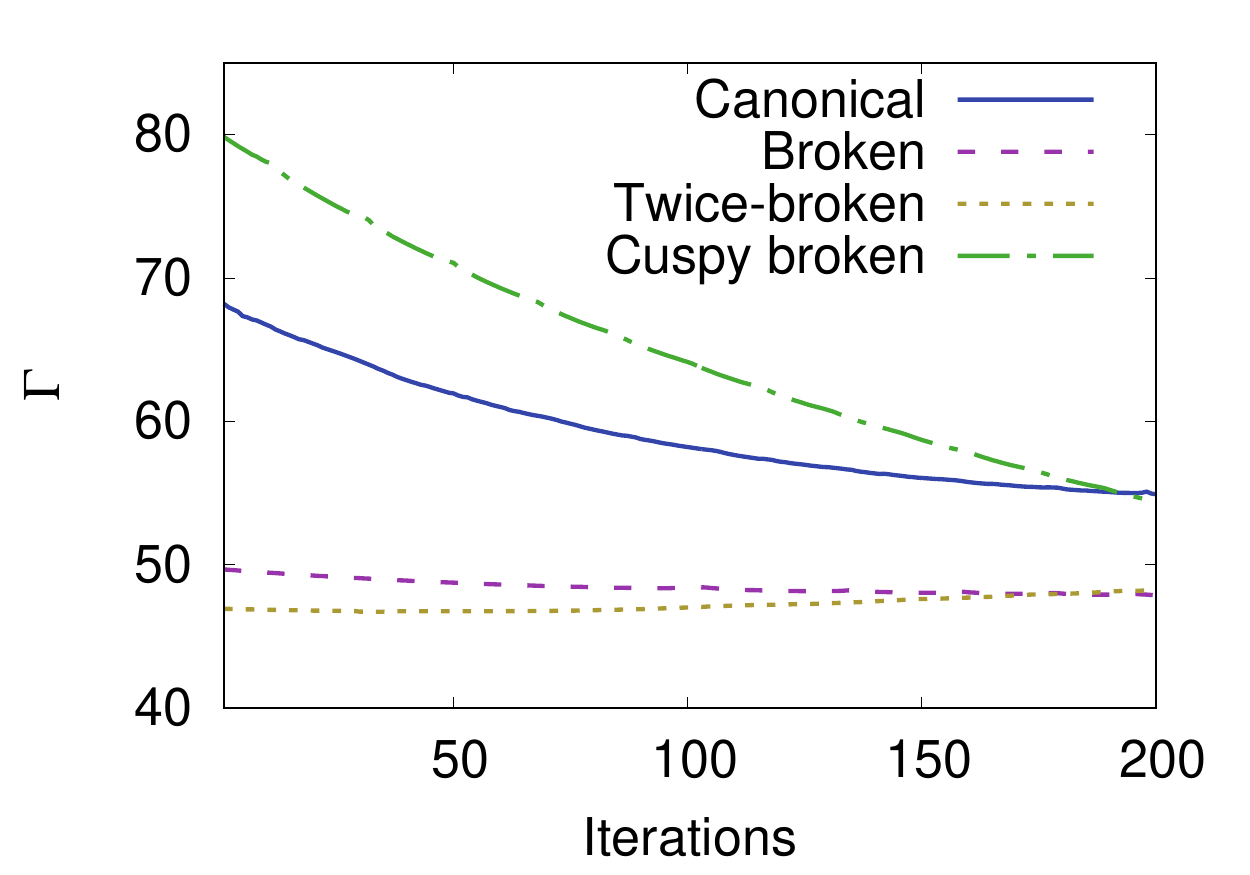}
    \caption{The changes in loop $\Gamma$
     due to gravitational backreaction.}\label{fig:cusps-gammas}
\end{figure}

In Fig.~\ref{fig:cusps-gammas},
we show the evolution of the $\Gamma$ factors for all loops discussed
above.  Loops that start with cusps have higher $\Gamma$, which is not
surprising.  Such loops preferentially lose energy from the region
around the cusp.  This leads to a decline in the cusp radiation and
contributes to a decline in the overall $\Gamma$.

Loops without cusps initially start with lower $\Gamma$, and there is
little change in $\Gamma$ over time.  Backreaction introduces cusps,
but the emission from them is always weaker than that of cusps present
initially.  The production of cusps does not increase the overall
$\Gamma$, so the (fairly small) emission from the cusps must be offset
by decreases in emission elsewhere.
 
We further observe that the changes to the $\Gamma$ values is in
a rough correspondence to the changes to the cusp strengths seen in Fig.~\ref{fig:cusp-strength}.

In the end, all loops appear to evolve towards a $\Gamma$ in the high
$40$s or low $50$s, although there does not appear to be a single
asymptotic value.  This is similar to the $\Gamma\approx50$ for loops
taken from simulations and smoothed~\cite{Blanco-Pillado:2015ana},
although this does not explain why a simple model loop would move
towards a configuration similar to a loop generated by a stochastic
process.

Note that the change in $\Gamma$ is due to a change in the shape of
the string, and not to its decreasing length. This change in shape
should also change the power spectrum, $P_n$, of the string,
particularly in the high-$n$ regime where the difference between kinks
and cusps dominates.

\subsection{Self-Intersections}

One of the important questions one would like to address is how robust non-self-intersecting trajectories are to the effects of backreaction. This was studied in detail for realistic loops obtained from a large scale simulation in \cite{Blanco-Pillado:2015ana} using a toy model for backreaction based on smoothing. This led to the conclusion that backreaction usually did not deform non-self-intersecting loops into self-intersecting trajectories. Here we revisit this issue, but with explicit backreaction in place of a toy model.

We check for self-intersections by taking the backreacted loops at
various points in their evolution and letting them undergo one full
oscillation in flat space. During this motion, we are sensitive to any
crossing of segments, which would lead to the loop fragmenting into
two child loops. We have not found any such self-intersections
in these cases. This seems to be the generic situation for this family
of loops.

\section{Conclusions}\label{sec:conc}

We have developed and demonstrated a technique for calculating
gravitational backreaction on cosmic string loops, although we have
only studied simple models in this work. This was done in order to
draw conclusions on the fates of cusps and kinks in as controlled of an
environment as possible. However, we are currently studying
backreaction on realistic cosmic string loops, as will be reported in
a future paper.

As expected from analytic work \cite{Blanco-Pillado:2018ael},
backreaction acting on one side of a kink rounds it off immediately,
but only over a narrow region of the string.  Viewed very close up,
the string is smooth, but at larger distances it still looks like a
kink.  Smoothing produces a cusp that was not there initially, but
this cusp is very weak, and never grows very strong as compared to
what one would expect for a loop whose  $\bA'$ and $\bB'$ move
uniformly around the unit sphere.

There are two reasons that cusps never grow very strong.  First, the
amount of the initial string involved in the rounding of the kink
grows only slowly with time.  But secondly, the energy in this string
is always being depleted, so that even as more and more of the initial
segments of string are involved in the cusp, the amount of energy in
each segment is going down.

For strings with cusps initially, the amount of energy involved in the
cusp, and consequently the cusp strength, declines over time by a 
factor of a few by the time the loop is about half evaporated.

Loops produced in simulations have many kinks, but no
cusps~\cite{Blanco-Pillado:2015ana}, because the paths of $\bA'$
and $\bB'$ often jump over each other but never cross smoothly.
Thus we expect the results on the initially cuspless loops that we
study in this paper to be the ones relevant for prediction of
observable signals from a cosmic string network.  These cusps never
grow to more than about one-tenth of the naive cusp strength that one
would predict for a smooth loop.  This worsens the prospects for
detection of burst signals, such as gravitational waves, coming from
cusps, and thus weakens the constraints from non-detection.

\begin{figure}
    \centering
    \includegraphics[scale=0.67]{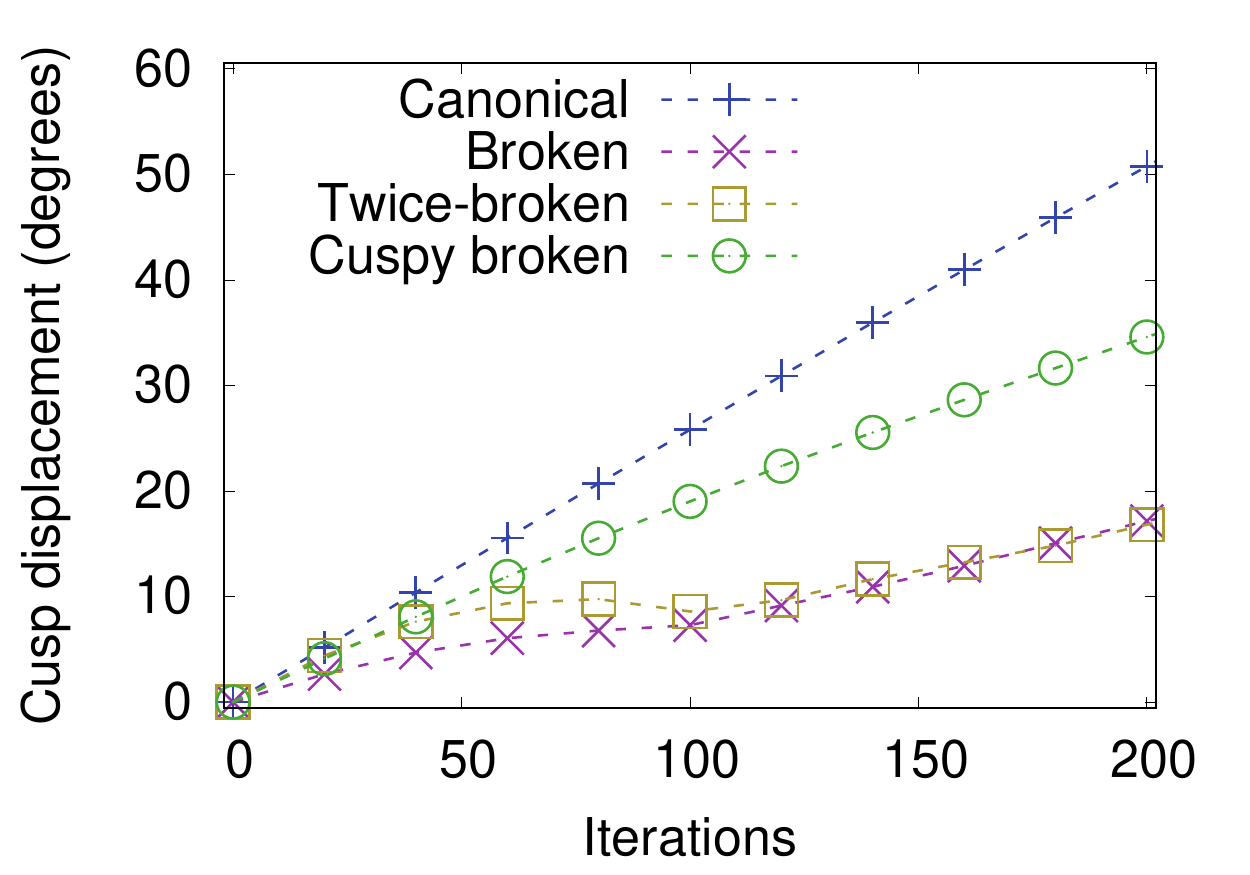}
    \caption{Motion of the cusp direction around the unit sphere due
      to backreaction.}\label{fig:cusp-motion}
\end{figure}

Cusps are ``dragged'' about the unit sphere in the general direction
of $\bX''$.  Thus successive bursts of gravitational waves from
cusps are emitted in slightly different directions, so one would not expect
observations of repeating bursts.  Figure~\ref{fig:cusp-motion} shows
the angular distance on the unit sphere between the direction of each cusp
and where it was originally.
In roughly the first half of the loop lifetime, studied here, the
angles are no more than $50^\circ$.  This has some impact on the
rocket effect~\cite{Vachaspati:1984gt,Durrer:1989zi} because the
thrust due to gravitational wave emission is not in a single
direction, but it is tiny. For $\theta = 50^\circ$, the average thrust
is only reduced 3\% over what it would be without dragging.\footnote{If the thrust is evenly distributed along
a great circle segment of angle $\theta$, the magnitude of the average
thrust is $(1/\theta) \int_{-\theta/2}^{\theta/2} d\phi\, \cos \phi =
(2/\theta) \sin (\theta/2)\approx 1-\theta^2/24$ if $\theta$ is
small.}

Finally, we note that describing the effect of backreaction on a loop
is not as simple as saying that cusps are weakened and kinks are
rounded off. These processes indeed take place, but parts of the
loop far from kinks or cusps are also affected, in complex ways.  In
particular we see that the dragging process affects segments near
kinks or cusps more than those further away, introducing features that
were not originally present.  The resulting loops are not simply
described as having, or not having, cusps and kinks.

\section*{Acknowledgments}\label{sec:ack}

We would like to thank David Chernoff and Alex Vilenkin for useful
conversations. The loop evolution computations and some figure
production were done on the Tufts Linux Research Cluster. This work
was supported in part by the National Science Foundation under grant
numbers 1518742, 1520792, and 1820902, the Spanish Ministry MINECO
grant (FPA2015-64041-C2-1P), and Basque Government grant
(IT-979-16). J. J. B.-P. is also supported in part by the Basque
Foundation for Science (IKERBASQUE).

\appendix

\section{Corrections to a segment due to a single source diamond}\label{app:segment-corrections}

Given an observer diamond and a source diamond on some string worldsheet, we may find the correction to the observer $A'$ and $B'$ due to that source. To do this, we make use of the $uvcd$ coordinates of Ref.~\cite{Blanco-Pillado:2018ael}. These are pseudo-orthonormal coordinates whose basis vectors are $B'/2$ for the $u$ direction and $A'/2$ for the $v$ direction, using the null vectors of the source, plus two spacelike vectors for the $c$ and $d$ direction which are orthogonal to the plane of the source diamond and to each other. Thus the source diamond is parameterized by the null parameters $u$ and $v$. This also means that
\begin{subequations}\label{eqn:AB-uvcd}\begin{align}
    A'^\gamma &= (0,2,0,0)\,,\\
    B'^\gamma &= (2,0,0,0)
\end{align} \end{subequations} 
when $\gamma=(u,v,c,d)$.

Some additional definitions will make the following equations more compact. First we say that the observer's motion in the observer diamond is along the null vector $V$, parameterized by $x$. Thus, if $\kappa$ connects the centers of the source and observer diamonds, then we can define
\begin{equation}
    \Omega(x) = \kappa + \frac{xV}{2}
\end{equation}
as the location of the observer relative to the source diamond's center, which we take as the origin of our coordinate system.  The edges of the source diamond have length $L_A$ and $L_B$, and so locally $u$ runs over $-L_B\ldots L_B$, and similarly for $v$ and $L_A$. This means that we can define vectors
\begin{subequations} \begin{align}
    E &= \Omega - L_AA'/2 - L_BB'/2\,,\\
    S &= \Omega + L_AA'/2 + L_BB'/2\,,
\end{align} \end{subequations}
which point from the future tip and past tip of the source diamond, respectively, to the observer. We also define $Z=A'\cdot B'$ for convenience, and indicate the null vectors of the observer diamond by $\bar A'$ and $\bar B'$ to distinguish them from the source diamond null vectors. Finally, we use the freedom in $c$ and $d$ to choose our coordinates such that $V^d=0$ always.

We will now take Eq.~(19) of Ref.~\cite{Wachter:2016rwc},
\begin{equation}
    h_{\alpha\beta} = \frac{8G\mu\sigma_{\alpha\beta}}{Z}\ln\left[\frac{2A'\Omega}{Z}-u\right]^{u_+}_{u_-}\,,
\end{equation}
with $h_{\alpha\beta}$ the first-order perturbation of the spacetime metric due to the source diamond,
\begin{equation}
    \sigma_{\alpha\beta}=\frac12\left(A'_\alpha B'_\beta+B'_\alpha A'_\beta - 4Z\eta_{\alpha\beta}\right)\,,
\end{equation}
$\eta_{\alpha\beta}$ the flat-space metric, and $u_\pm$ the maximum and minimum values of that parameter visited by the intersection line within the source diamond. Then, making use of Eqs.~(\ref{eqn:xab},\ref{eqn:delta-a'b'}), we find that the correction to a null vector $\bar N'$ of the observer diamond due to the source diamond is given by
\begin{align}\label{eqn:Delta-N'}
    \Delta\bar N^\lambda &= -G\mu\left[\left(\bar A'^v A'^\lambda + \bar A'^u B'^\lambda - 2\bar A'^\lambda\right)F_\rho\bar B'^\rho+\left(\bar B'^v A'^\lambda + \bar B'^u B'^\lambda - 2\bar B'^\lambda\right)F_\rho\bar A'^\rho\right.\nonumber\\
    &\qquad\qquad\left.+F^\lambda\left(\bar A'^c\bar B'^c + \bar A'^d\bar B'^d\right)\right]\,,
\end{align}
where $F$ is given by
\begin{equation}
    \int^{x_f}_{x_i} h_{\alpha\beta,\gamma} = \frac{8G\mu\sigma_{\alpha\beta}}{Z}F_\gamma\,.
\end{equation}
The values of $F_\gamma$ depend on which of the three types of crossing discussed in Sec.~\ref{ssec:changes} the source diamond possesses. This crossing type may change as we move along the observer line in the observer diamond, and thus a single source diamond could contribute up to three separate $\Delta \bar N'$ terms which correct the observer diamond's null vector. The initial and final values of the null parameter we integrate over, $x_i$ and $x_f$, give the range of the observer's motion for a given crossing type.

For an intersection line which connects opposite edges of fixed $u$,
\begin{subequations} \begin{align}
    F_u &= \frac{2}{V^u}\left(\ln\left[\frac{E^u(x_f)}{E^u(x_i)}\right]-\ln\left[\frac{S^u(x_f)}{S^u(x_i)}\right]\right)\,,\\
    F_v &= 0\,,\\
    F_c &= 0\,,\\
    F_d &= 0\,.
\end{align} \end{subequations}
For an intersection line which connects opposite edges of fixed $v$, the $F$ are the same as the case which connects edges of fixed $u$, but with $u\leftrightarrow v$. For an intersection line which connects the two future edges of the source diamond,
\begin{subequations} \begin{align}
    F_u &= \frac{2}{V^u}\ln\left[\frac{E^u(x_f)}{E^u(x_i)}\right]\,,\\
    F_v &= \frac{2}{V^v}\ln\left[\frac{E^v(x_f)}{E^v(x_i)}\right]\,,\\
    F_c &= -\frac{2}{V^c}\ln\left[\frac{(\Omega^c(x_f))^2+(\Omega^d)^2}{(\Omega^c(x_i))^2+(\Omega^d)^2}\right]\,,\\
    F_d &= -\frac{4}{V^c}\arctan\left[\frac{(\Omega^c(x_f)-\Omega^c(x_i))\Omega^d}{(\Omega^d)^2+\Omega^c(x_f)\Omega^c(x_i)}\right]\,.
\end{align} \end{subequations}
Note that because $V^d=0$, $\Omega^d$ has no dependence on $x$. Finally, for an intersection line which connects the two past edges of the source diamond, the $F$ are as the case which connects the two future edges, but with the overall sign of each $F$ changed and with $E\rightarrow S$.

With the forms of the $F$, and Eq.~(\ref{eqn:AB-uvcd}), we may simplify Eq.~(\ref{eqn:Delta-N'}). For a $u$-type crossing, we know that only $F_u\neq 0$, and so the velocity correction becomes
\begin{subequations}\label{eqn:d-vel-u}\begin{align}
    \Delta \bar N^u &= 0\,,\\
    \Delta \bar N^v &= -\frac{G\mu F_u}{Z}\left(\bar A'^c\bar B'^c + \bar A'^d\bar B'^d\right)\,,\\
    \Delta \bar N^c &= G\mu F_u \left(\bar B'^u\bar A'^c+\bar A'^u\bar B'^c\right)\,,\\
    \Delta \bar N^d &= G\mu F_u \left(\bar B'^u\bar A'^d+\bar A'^u\bar B'^d\right)\,.
\end{align}\end{subequations}
For a $v$-type crossing, by the usual symmetry of $u\leftrightarrow v$ and $A'\leftrightarrow B'$, we find
\begin{subequations}\label{eqn:d-vel-v}\begin{align}
    \Delta \bar N^u &= -\frac{G\mu F_v}{Z}\left(\bar A'^c\bar B'^c + \bar A'^d\bar B'^d\right)\,,\\
    \Delta \bar N^v &= 0\,,\\
    \Delta \bar N^c &= G\mu F_v \left(\bar B'^v\bar A'^c+\bar A'^v\bar B'^c\right)\,,\\
    \Delta \bar N^d &= G\mu F_v \left(\bar B'^v\bar A'^d+\bar A'^v\bar B'^d\right)\,.
\end{align}\end{subequations}
For a past- or future-type crossing, no member of $F_\gamma$ is generally zero. So,
\begin{subequations}\label{eqn:d-vel-past}\begin{align}
    \Delta \bar N^u &= -\frac{G\mu F_v}{Z}\left(\bar A'^c\bar B'^c + \bar A'^d\bar B'^d\right)\,,\\
    \Delta \bar N^v &= -\frac{G\mu F_u}{Z}\left(\bar A'^c\bar B'^c + \bar A'^d\bar B'^d\right)\,,\\
    \Delta \bar N^c &= G\mu\bar A'^c F_\rho\bar B'^\rho + G\mu\bar B'^c F_\rho\bar A'^\rho - G\mu F_c\left(\bar A'^c\bar B'^c + \bar A'^d\bar B'^d\right)\,,\\
    \Delta \bar N^d &= G\mu\bar A'^d F_\rho\bar B'^\rho + G\mu\bar B'^d F_\rho\bar A'^\rho - G\mu F_d\left(\bar A'^c\bar B'^c + \bar A'^d\bar B'^d\right)\,,
\end{align}\end{subequations}
with the difference in the two crossing types coming entirely from the $F_\gamma$ terms.

\section{Calculating $\Gammac$}\label{app:gamma-cusp}

The angular power density in gravitational waves emitted by a cusp
diverges as the observer approaches the cusp direction.  We would like
to use the coefficient of this divergence to characterize the strength
of the cusp.

We begin by considering a coordinate system oriented so that
$\bA'=\bB'$ points entirely in the $z$ direction, so $\bA''$ and
$\bB''$ lie entirely in the $x$-$y$ plane. We establish spherical
polar coordinates $(\theta,\phi)$, where $\theta=0$ is the cusp
direction.  Let $(\tobs,\pobs)$ denote the direction of the observer
in these coordinates.  We consider directions close to the cusp,
$\tobs\ll1$.  The directions of $\bA''$ and $\bB''$ are $(\pi/2,\phi_A)$
and $(\pi/2,\phi_B)$ respectively.  Define the relative angles $\phi_{AO}
= \phi_A-\phi_O$ and $\phi_{BO} = \phi_B-\phi_O$, and $\phi_{AB} =
\phi_A-\phi_B$.

The power per unit frequency $\omega$ per unit solid angle is given by
Eq.~(A29) of Ref.~\cite{Blanco-Pillado:2017oxo},
\begin{align}
    \frac{dP}{d\omega d\Omega} =& \frac{2G\mu^2\omega^2\tobs^8}{9\pi^2L}\frac{\sin^4\phi_{AO}\sin^4\phi_{BO}}{|\bA''|^2|\bB''|^2}\left[(K^2_{1/3}(\xi_A)+K^2_{2/3}(\xi_A))(K^2_{1/3}(\xi_B)+K^2_{2/3}(\xi_B))\right.\nonumber\\
    &\left.+4\operatorname{sign}(\sin\phi_{AO}\sin\phi_{BO})K_{1/3}(\xi_A)K_{2/3}(\xi_A)K_{1/3}(\xi_B)K_{2/3}(\xi_B)\right]\,.
\end{align}
Here $K_\alpha$ is the modified Bessel function of the second kind,
and we have defined
\begin{equation}
    \xi_A = \omega\tobs^3\frac{|\sin^3\phi_{AO}|}{6|\bA''|}
\end{equation}
and likewise for $\xi_B$.

We change variables from $\omega$ to
\begin{equation}
w= \sqrt{\frac{|\sin^3\phi_{AO}\sin^3\phi_{BO}|}{36|\bA''||\bB''|}}\tobs^3\omega
\end{equation}
and integrate over $w$ to get
\begin{equation}\label{eqn:dPdO}
    \frac{dP}{d\Omega} = \frac{48G\mu^2}{\tobs\pi^2L}\frac{1}{(|\bA''||\bB''||\sin\phi_{AO}\sin\phi_{BO}|)^{1/2}}\mathcal{H}_{\sign(\sin\phi_{AO}\sin\phi_{BO})}(a)\,,
\end{equation}
where
\begin{equation}
    a = \frac{\xi_A}{\xi_B} = \frac{|\bB''|}{|\bA''|}\left|\frac{\sin\phi_{AO}}{\sin\phi_{BO}}\right|^3
\end{equation}
and
\begin{align}\label{eqn:Ha}
    \mathcal{H}_\pm(a) = \int^\infty_0 w^2&\left[(K^2_{1/3}(a^{-1/2}w)+K^2_{2/3}(a^{-1/2}w))(K^2_{1/3}(a^{1/2}w)+K^2_{2/3}(a^{1/2}w))\right.\nonumber\\
    &\left.\pm4K_{1/3}(a^{-1/2}w)K_{2/3}(a^{-1/2}w)K_{1/3}(a^{1/2}w)K_{2/3}(a^{1/2}w)\right]dw\,.
\end{align}
This is invariant under $a\rightarrow1/a$, and thus Eq.~(\ref{eqn:dPdO}) is invariant under the interchange of $A$ and $B$. It is also invariant under rescalings of the loop length, as both $|\bA''|$ and $|\bB''|$ go like $1/L$. Length-invariance makes this quantity a good measure of cusp strength for considering how backreaction changes a cusp on a loop over time, as the loop's length is also changing due to backreaction.

In the main text we defined $\Gamma_\text{cusp}(\theta)$ to be the
contribution to $\Gamma$ coming from angles within $\theta$ of the
cusp direction.  So we should compute
\begin{equation}
     \int_0^{2\pi}d\pobs \int_0^\theta \sin\tobs\, d\tobs\,  \frac{dP}{d\Omega}
\end{equation}
and then use $P = G\mu^2\Gamma$ to find the contribution to $\Gamma$.
The polar integration is straightforward, because we are working in
the regime where $\tobs\ll1$ and thus $\sin\tobs\approx\tobs$. The
$\tobs$ here cancels the $\tobs$ in the denominator of
Eq.~(\ref{eqn:dPdO}), and so our expression is overall
$\propto\theta$.  Due to the dependence of $a$ and $w$ on $\pobs$, the
azimuthal integration must be done numerically.  This integration
gives some number $\Gammac/\theta$, which we show in
Fig.~\ref{fig:cusp-strength}.

\bibliography{paper}

\end{document}